\documentclass[12pt,thmsa]{article}
\usepackage{amsmath}

\usepackage{amsfonts}
\usepackage{sw20mitp}

\input{tcilatex}

\begin{document}

\author{Jean PETITOT\thanks{%
CAMS, Ecole des Hautes Etudes en Sciences Sociales, Paris.
petitot@poly.polytechnique.fr}}
\title{Noncommutative geometry and physics
(February 2007)}

\date{}

\maketitle

\noindent This is a compilation of some well known propositions of Alain
Connes concerning the use of noncommutative geometry in mathematical
physics. It has been used in my paper \textquotedblleft Noncommutative
Geometry and Transcendental Physics\textquotedblright\ in \emph{Constituting
Objectivity. Transcendental Perspectives on Modern Physics}, (M. Bitbol, P.
Kerszberg, J. Petitot, eds), Springer, 2009.

\bigskip

\bigskip

\bigskip

\tableofcontents

\newpage

\section{Gelfand theory}

To understand noncommutative geometry we must first come back to Gelfand
theory for \emph{commutative} $C^{\ast }$-algebras.

\subsection{$C^{*}$-algebras}

Recall that a $C^{\ast }$-algebra $\mathcal{A}$ is a (unital) Banach algebra
on $\mathbb{C}$ (i.e. a $\mathbb{C}$-algebra which is normed and complete
for its norm) endowed with an involution $x\rightarrow x^{\ast }$ s.t. $%
\left\Vert x\right\Vert ^{2}=\left\Vert x^{\ast }x\right\Vert $. The norm
(the metric structure) is then deducible from the algebraic structure.$\;$%
Indeed, $\left\Vert x\right\Vert ^{2}$ is the spectral radius of the $\geq 0$
element $x^{\ast }x$, that is, the $\func{Sup}$ of the modulus of the
spectral values of $x^{\ast }x$:\footnote{%
In the infinite dimensional case, the spectral values ($x-\lambda I$ is not
invertible) are not identical with the eigenvalues ($x-\lambda I$ has a non
trivial kernel). Indeed non invertibility no longer implies non injectivity
(a linear operator can be injective and non surjective). For instance, if $%
e_{n}$, $n\in \mathbb{N}$, is a countable basis, the shift $\sum_{n}\lambda
_{n}e_{n}\rightarrow \sum_{n}\lambda _{n}e_{n+1}$ is injective but not
surjective and is not invertible.}$\;$%
\[
\left\Vert x\right\Vert ^{2}=\func{Sup}\left\{ \left\vert \lambda
\right\vert :x^{\ast }x-\lambda I\text{ is not invertible}\right\} 
\]%
(where $I$ is the unit of $\mathcal{A}$).\ In a $C^{\ast }$-algebra the norm
becomes therefore a purely \emph{spectral} concept.

An element $x\in \mathcal{A}$ is called self-adjoint if $x=x^{*}$, normal if 
$xx^{*}=x^{*}x$, and unitary if $x^{-1}=x^{*}$ ($\left\| x\right\| =1$).

In this classical setting, the mathematical interpretations of fundamental
physical concepts such as a space of states, an observable, or a measure,
are the following:

\begin{enumerate}
\item a space of states is a smooth manifold: the phase space $M$ (in
Hamiltonian mechanics, $M=T^{\ast }N$ is the cotangent bundle of the space
of configurations $N$ endowed with its canonical symplectic structure);

\item an observable is a function $f:M\rightarrow \mathbb{R}$ (interpreted
as $f:M\rightarrow \mathbb{C}$ with $f=\bar{f}$) which measure some property
of states and output a real number;

\item the measure of $f$ in the state $x\in M$ is the evaluation $f(x)$ of $f
$ at $x$; but as $f(x)=\delta _{x}(f)$ (where $\delta _{x}$ is the Dirac
distribution at $x$) a state can be dually interpreted as a continuous
linear operator on observables.
\end{enumerate}

Observables constitute a commutative $C^{\ast }$-algebra $\mathcal{A}$ and
Gelfand theory explains that the \emph{geometry} of the manifold $M$ can be
completely recovered from the \emph{algebraic} structure of $\mathcal{A}$.

\subsection{Gelfand's theorem}

Let $M$ be a topological space and let $\mathcal{A}:=\mathcal{C}(M)$ be the $%
\mathbb{C}$-algebra of continuous functions $f:M\rightarrow \mathbb{C}$ (the 
$\mathbb{C}$-algebra structure being inherited from the structure of $%
\mathbb{C}$ via pointwise addition and multiplication).\ Under very general
conditions (e.g. if $M$ is compact~\footnote{%
If $M$ is non compact but only locally compact, then one take $\mathcal{A}=%
\mathcal{C}_{0}(M)$ the algebra of continuous functions vanishing at
infinity but $\mathcal{A}$ is no longer unital since the constant function $1
$ doesn't vanish at infinity.}), it is a $C^{\ast }$-algebra for complex
conjugation $f^{\ast }=\overline{f}$.

The possible values of $f$~-- that is the possible results of a measure of $f
$~-- can be defined in a purely algebraic way as the \emph{spectrum} of $f$
that is 
\[
\limfunc{sp}\nolimits_{\mathcal{A}}(f):=\left\{ c:f-cI\text{ is not
invertible in }\mathcal{A}\right\} ~.
\]

\noindent Indeed, if $f(x)=c$ then $f-cI$ is not invertible in $\mathcal{A}$%
. $\limfunc{sp}\nolimits_{\mathcal{A}}(f)$ is the complementary set of what
is called the \emph{resolvent} of $f$, 
\[
r(f):=\left\{ c:f-cI\text{ is invertible in }\mathcal{A}\right\} ~.
\]

The main point is that the evaluation process $f(x)$~-- that is measure~--
can be interpreted as a \emph{duality} $\left\langle f,x\right\rangle $
between the space $M$ and the algebra $\mathcal{A}$. Indeed, to a point $x$
of $M$ we can associate the \emph{maximal ideal} of the $f\in \mathcal{A}$
vanishing at $x$: 
\[
\mathfrak{M}_{x}:=\left\{ f\in \mathcal{A}:f(x)=0\right\} ~.
\]

\noindent But the maximal ideals $\mathfrak{M}$ of $\mathcal{A}$ constitute
themselves a space~-- called the \emph{spectrum} of the algebra $\mathcal{A}$%
. They can be considered as the kernels of the \emph{characters} of $%
\mathcal{A}$, that is of the morphisms (multiplicative linear forms) $\chi :%
\mathcal{A}\rightarrow \mathbb{C}$, 
\[
\mathfrak{M=}\chi ^{-1}(0)~.
\]

\noindent A character is by definition a coherent procedure for evaluating
together all the elements $f\in \mathcal{A}$. The evaluation $\chi (f)$ is
also a \emph{duality} $\left\langle \chi ,f\right\rangle $.and its outputs $%
\chi (f)$ belong to $\limfunc{sp}\nolimits_{\mathcal{A}}(f)$.\ Indeed, as 
\emph{distributions} (continuous linear forms), the characters correspond to
the Dirac distributions $\delta _{x}$ and if $\chi =\delta _{x}$, then $\chi
(f)=f(x)=c$ and $c\in \limfunc{sp}\nolimits_{\mathcal{A}}(f)$.

The \emph{spectrum} of the $C^{\ast }$-algebra $\mathcal{A}$ (not to be
confused with the spectra $\limfunc{sp}\nolimits_{\mathcal{A}}(f)$ of the
single elements $f$ of $\mathcal{A}$) is by definition the space of
characters $\func{Sp}(\mathcal{A}):=\{\chi \}$ endowed with the topology of
simple convergence: $\chi _{n}\rightarrow \chi $ iff $\chi
_{n}(f)\rightarrow \chi (f)$ for every $f\in \mathcal{A}$. It is defined
uniquely from $\mathcal{A}$ without any reference to the fact that $\mathcal{%
A}$ is of the form $\mathcal{A}:=\mathcal{C}(M)$. It is also the space of
irreducible representations of $\mathcal{A}$ (since $\mathcal{A}$ is
commutative, they are $1$-dimensional).

Now, if $f\in \mathcal{A}$ is an element of $\mathcal{A}$, using duality, we
can associate to it canonically a \emph{function} $\tilde{f}$ on the space $%
\func{Sp}(\mathcal{A})$%
\[
\begin{array}{rll}
\tilde{f}:\func{Sp}(\mathcal{A}) & \rightarrow  & \mathbb{C} \\ 
\chi  & \mapsto  & \tilde{f}(\chi )=\chi (f)=\left\langle \chi
,f\right\rangle ~.%
\end{array}%
\]

\noindent We get that way a map 
\[
\begin{array}{rll}
\symbol{126}:\mathcal{A} & \rightarrow & \mathcal{C(}\func{Sp}\mathcal{(%
\mathcal{A}))} \\ 
f & \mapsto & \tilde{f}%
\end{array}
\]

\noindent which is called the \emph{Gelfand transform}. For every $f$ we
have 
\[
\tilde{f}\left( \func{Sp}(\mathcal{A})\right) =\limfunc{sp}\nolimits_{%
\mathcal{A}}(f)~.
\]

The key result is then:

\textbf{Gelfand-Neimark theorem}.\ If $\mathcal{A}$ is a \emph{commutative} $%
C^{*}$-algebra, the Gelfand transform $\symbol{126}$ is an \emph{isometry}
between $\mathcal{A}$ and $\mathcal{C(}\func{Sp}\mathcal{(\mathcal{A}))}$.

Indeed, the norm of $\tilde{f}$ is the spectral radius of $f$, $\rho \left(
f\right) :=\underset{n\rightarrow \infty }{\lim }\left( \left\|
f^{n}\right\| ^{\frac{1}{n}}\right) $ and we have $\left\| \tilde{f}\right\|
=\rho \left( f\right) =\left\| f\right\| $.\ To see this, suppose first that 
$f$ is self-adjoint ($f=f^{*}=\bar{f}$).\ We have $\left\| f\right\|
^{2}=\left\| f.f^{*}\right\| =\left\| f^{2}\right\| $.\ So, $\left\|
f\right\| =\left\| f^{2^{n}}\right\| ^{2^{-n}}$ and as $\left\|
f^{2^{n}}\right\| ^{2^{-n}}\rightarrow \rho \left( f\right) $ by definition
we have $\left\| f\right\| =\rho \left( f\right) $.\ Suppose now that $f$ is
any element of $\mathcal{A}$. Since $f.f^{*}$ is self-adjoint, we have $%
\left\| f\right\| ^{2}=\left\| f.f^{*}\right\| =\rho \left( f.f^{*}\right)
=\left\| \widetilde{f.f^{*}}\right\| $.\ But$\left\| \widetilde{f.f^{*}}%
\right\| =\left\| \widetilde{f}.\widetilde{f^{*}}\right\| =\left\| 
\widetilde{f}\right\| ^{2}$ and therefore $\left\| f\right\| ^{2}=\left\| 
\widetilde{f}\right\| ^{2}$ and $\left\| f\right\| =\left\| \widetilde{f}%
\right\| $.

Gelfand theory shows that, in the classical case of commutative $C^{*}$%
-algebras $\mathcal{A}:=\mathcal{C}(M)$ ($M$ compact), there exists a
complete \emph{equivalence} between the geometric and the algebraic
perspectives.\ 

\subsection{Towards a new kinematics}

We think that Gelfand theorem has a deep philosophical meaning. In classical
mechanics kinematics concerns the structure of the configuration spaces $N$
and phase spaces $M:=T^{\ast }N$, and motions and trajectories in them.
Observables and measurements are defined in terms of functions on these
basic spaces directly construed from the geometry of space-time. Gelfand
theorem shows than we can \emph{exchange} the primary geometrical background
with the secondary process of measure, take measure as a primitive fact and
reconstruct the geometric background from it.

\subsection{Towards Noncommutative Geometry}

In Quantum Mechanics, the basic structure is that of the \emph{noncommutative%
} $C^{\ast }$-algebras $\mathcal{A}$ of observables. It is therefore natural
to wonder if there could exist a \emph{geometric }correlate of this
noncommutative algebraic setting.\ It is the origin of Connes'
Noncommutative Geometry (NCG) also called Spectral Geometry or Quantum
Geometry. In NCG the basic structure is the NC $C^{\ast }$-algebra $\mathcal{%
A}$ of obervables: any phenomenon is something which is observable in the
quantum sense, and not an event in space-time. But observables must be
defined for states and are therefore represented in the space of states of
the system, which is an Hilbert space and not the classical space. The
associated NC space is then the space of irreducible representations of $%
\mathcal{A}$.

NCG is a fundamentally new step toward a geometrization of physics. Instead
of beginning with classical differential geometry and trying to develop
Quantum Mechanics on this backgrond, it begins with Quantum Mechanics and
construct a new quantum geometrical framework.\ The most fascinating aspect
of Connes' research program is how he succeeded in reinterpreting all the
basic structures of classical geometry inside the framework of NC $C^{\ast }$%
-algebras operating on Hilbert spaces. The basic concepts remain almost the
same but their mathematical interpretation is significantly complexified,
since their classical meaning becomes a \emph{commutative limit}. We meet
here a new very deep example of the conceptual transformation of physical
theories through mathematical enlargements, as it is the case in general
relativity. As explained by Daniel Kastler \cite{KastlerNCG}:

\begin{quotation}
\noindent 
``Alain Connes' noncommutative geometry (...) is a systematic quantization
of mathematics parallel to the quantization of physics effected in the
twenties. (...) This theory widens the scope of mathematics in a manner
congenial to physics.''
\end{quotation}

\section{NCG and differential forms}

Connes reinterpreted (in an extremely deep and technical way) the six
classical levels:

\begin{enumerate}
\item Measure theory;

\item Algebraic topology and topology ($K$-theory);

\item Differentiable structure;

\item Differential forms and De Rham cohomology;

\item Fiber bundles, connections, covariant derivations, Yang-Mills theories;

\item Riemannian manifolds and metric structures.
\end{enumerate}

Let us take as a first example the reinterpretation of the differential
calculus.

\subsection{A universal and formal differential calculus}

How can one interpret differential calculus in the new NC paradigm? One
wants first to define \emph{derivations} $D:\mathcal{A}\rightarrow \mathcal{E%
}$, that is $\mathbb{C}$-linear maps satisfying the \emph{Leibniz rule}
(which is the universal formal rule for derivations): 
\[
D(ab)=(Da)b+a(Db)~.
\]

\noindent For that, $\mathcal{E}$ must be endowed with a structure of $%
\mathcal{A}$-bimodule (right and left products of elements of $\mathcal{E}$
by elements of $\mathcal{A}$).\ It is evident that $D(c)=0$ for any scalar $%
c\in \mathbb{C}$ since $D(1.a)=D(1)a+1D(a)=D(a)$ and therefore $D(1)=0$.

Let $\func{Der}(\mathcal{A},\mathcal{E})$ be the $\mathbb{C}$-vector space
of such derivations. In $\func{Der}(\mathcal{A},\mathcal{E})$ there exist
very particular elements, the \emph{inner} derivatives, associated with the
elements $m$ of $\mathcal{E}$, which express the difference between the
right and left $\mathcal{A}$-module structures of $\mathcal{E}$: 
\[
D(a):=\func{ad}(m)(a)=ma-am~.
\]

\noindent Indeed, 
\begin{eqnarray*}
\func{ad}(m)(a).b+a.\func{ad}(m)(b) &=&(ma-am)b+a(mb-bm) \\
&=&mab-abm \\
&=&\func{ad}(m)(ab)~.
\end{eqnarray*}

\noindent In the case where $\mathcal{E}=\mathcal{A}$, $\func{ad}(b)(a)=%
\left[ b,a\right] $ expresses the non commutativity of $\mathcal{A}$. By the
way, $\func{Der}(\mathcal{A},\mathcal{A})$ is a Lie algebra since $\left[
D_{1},D_{2}\right] $ is a derivation if $D_{1},D_{2}$ are derivations.

Now, it must be stressed that there exists \emph{a universal derivation}
depending only upon the algebraic structure of $\mathcal{A}$ (supposed to be
unital), and having therefore nothing to do with the classical
\textquotedblleft infinitesimal\textquotedblright\ intuitions underlying the
classical concepts of differential and derivation. It is given by 
\[
\begin{array}{rll}
d:\mathcal{A} & \rightarrow  & \mathcal{A\otimes }_{\mathbb{C}}\mathcal{A}
\\ 
a & \mapsto  & da:=1\otimes a-a\otimes 1~.%
\end{array}%
\]

Let $\Omega ^{1}\mathcal{A}$ be the sub-bimodule of $\mathcal{A\otimes }_{%
\mathbb{C}}\mathcal{A}$ generated by the elements $adb:=a\otimes b-ab\otimes
1$, i.e. the kernel of the multiplication $a\otimes b\mapsto ab$.\footnote{%
For $a\otimes b-ab\otimes 1$ the multiplication gives $ab-ab=0$. Conversely
if $ab=0$ then $a\otimes b=a\otimes b-ab\otimes 1$ and $a\otimes b$ belongs
to $\Omega ^{1}\mathcal{A}$.} $\Omega ^{1}\mathcal{A}$ is isomorphic to the
tensorial product $\mathcal{A\otimes }_{\mathbb{C}}\overline{\mathcal{A}%
\text{,}}$ where $\overline{\mathcal{A}}$ is the quotient $\mathcal{A}/%
\mathbb{C}$ (i.e. $\mathcal{A}=\mathbb{C}1\oplus \overline{\mathcal{A}}$),
with $adb=a\otimes \overline{b}$. It is called the bimodule of \emph{%
universal }$1$\emph{-forms} on $\mathcal{A}$ where \textquotedblleft
universality\textquotedblright\ means that 
\[
\func{Der}(\mathcal{A},\mathcal{E})\simeq \limfunc{Hom}\nolimits_{\mathcal{A}%
}\left( \Omega ^{1}\mathcal{A},\mathcal{E}\right) 
\]

\noindent i.e. that a derivation $D:\mathcal{A}\rightarrow \mathcal{E}$ is
the same thing as a morphism of algebras between $\Omega ^{1}\mathcal{A}$
and $\mathcal{E}$. If $D:\mathcal{A}\rightarrow \mathcal{E}$ is an element
of $(\mathcal{A},\mathcal{E})$, the associated morphism $\tilde{D}:\Omega
^{1}\mathcal{A}\rightarrow \mathcal{E}$ is defined by 
\[
a\otimes b\mapsto aD(b)~.
\]

\noindent So $da=1\otimes a-a\otimes 1\mapsto 1.D(a)-a.D(1)=D(a)$ (since $%
D(1)=0$).

We can generalize this construction to universal $n$-forms, which have the
symbolic form\footnote{$da_{1}...da_{n}$ is the exterior product of $1$%
-forms, classically denoted $da_{1}\wedge ...\wedge da_{n}$.} 
\[
a_{0}da_{1}...da_{n}~.
\]

\noindent If $\Omega ^{n}\mathcal{A}:=\left( \Omega ^{1}\mathcal{A}\right)
^{\otimes _{\mathcal{A}}n}$ with $a_{0}da_{1}...da_{n}=a_{0}\otimes 
\overline{a_{1}}\otimes ...\otimes \overline{a_{n}}$, the differential is
then 
\[
\begin{array}{rll}
d:\Omega ^{n}\mathcal{A} & \rightarrow  & \Omega ^{n+1}\mathcal{A} \\ 
a_{0}da_{1}...da_{n} & \mapsto  & da_{0}da_{1}...da_{n} \\ 
a_{0}\otimes \overline{a_{1}}\otimes ...\otimes \overline{a_{n}} & \mapsto 
& 1\otimes \overline{a_{0}}\otimes \overline{a_{1}}\otimes ...\otimes 
\overline{a_{n}}~.%
\end{array}%
\]

\noindent Since $d1=0$, it is easy to verify the fundamental cohomological
property $d^{2}=0$ of the graduate differential algebra $\Omega \mathcal{A}%
:=\bigoplus_{n\in \mathbb{N}}\Omega ^{n}\mathcal{A}$. Some technical
difficulties must be overcome (existence of ``junk'' forms) to transform
this framework into a ``good'' formal differential calculus.

\subsection{Noncommutative differential calculus or ``quantized'' calculus}

To use this noncommutative differential in physics, Connes wanted to \emph{%
represent} the universal differential algebra in spaces of physical states.
Let us suppose therefore that the $C^{\ast }$-algebra $\mathcal{A}$ acts
upon an Hilbert space of states $\mathcal{H}$. One wants to interpret in
this representation the universal, formal, and purely symbolic differential
calculus of the previous section. For achieving that, one must interpret the
differential $df$ of the elements $f\in \mathcal{A}$ when these $f$ are
represented as \emph{operators} on $\mathcal{H}$. Connes' main idea was to
use the well-known formula of quantum mechanics 
\[
\frac{df}{dt}=\frac{2i\pi }{h}[F,f]
\]

\noindent where $F$ is the Hamiltonian of the system and $f$ any obervable.

Consequently, he interpreted the symbol $df$ as 
\[
df:=\left[ F,f\right] 
\]

\noindent for an appropriate self-adjoint operator $F$. One wants of course $%
d^{2}f=0.$\ But $d^{2}f=\left[ F^{2},f\right] $ and therefore $F^{2}$ must
commute with all observables.\ 

The main constraint is that, once interpreted in $\mathcal{H}$, the symbol $%
df$ must correspond to an \emph{infinitesimal}.\ The classical concept of
infinitesimal ought to be reinterpreted in the noncommutative framework.
Connes' definition is that an operator $T$ is infinitesimal if it is \emph{%
compact}, that is if the eigenvalues $\mu _{n}(T)$ of its absolute value $%
\left\vert T\right\vert =\left( T^{\ast }T\right) ^{1/2}$~-- called the 
\emph{characteristic values} of $T$~-- converge to $0$, that is if for every 
$\varepsilon >0$ the norm $\left\Vert T\right\Vert $ of $T$ is $<\varepsilon 
$ outside a subspace of\emph{\ finite} dimension. If $\mu _{n}(T)\underset{%
n\rightarrow \infty }{\rightarrow }0$ as $\frac{1}{n^{\alpha }}$ then $T$ is
an infinitesimal of order $\alpha $ ($\alpha $ is not necessarily an
integer). If $T$ is compact, let $\xi _{n}$ be a complete orthonormal basis
of $\mathcal{H}$ associated to $\left\vert T\right\vert $, $T=U\left\vert
T\right\vert $ the polar decomposition of $T$~\footnote{%
The polar decomposition $T=U\left\vert T\right\vert $ is the equivalent for
operators of the decomposition $z=\left\vert z\right\vert e^{i\theta }$ for
complex numbers. In general $U$ cannot be unitary but only a partial
isometry.} and $\eta _{n}=U\xi _{n}$.\ Then $T$ is the sum 
\[
T=\sum_{n\geq 0}\mu _{n}(T)\left\vert \eta _{n}\right\rangle \left\langle
\xi _{n}\right\vert ~.
\]

If $T$ is a positive infinitesimal of order $1$, its trace $\func{Trace}%
\left( T\right) =\sum_{n}\mu _{n}(T)$ has a logarithmic divergence.\ If $T$
is of order $>1$, its trace is finite $>0$. It is the basis for
noncommutative\ integration which uses the \emph{Dixmier trace}, a technical
tool for constructing a new trace extracting the logarithmic divergence of
the classical trace. Dixmier trace is a trick giving a meaning to the
formula $\underset{N\rightarrow \infty }{\lim }\frac{1}{\ln N}%
\sum_{n=0}^{n=N-1}\mu _{n}(T)$. It vanishes for infinitesimals of order $>1$.

Therefore, we interpret the differential calculus in the noncommutative
framework through triples $(\mathcal{A},\mathcal{H},F)$ where $\left[ F,f%
\right] $ is compact for every $f\in \mathcal{A}$.\ Such a structure is
called \emph{a Fredholm module}. The differential forms $a_{0}da_{1}...da_{n}
$ can now be interpreted as operators on $\mathcal{H}$ according to the
formula%
\[
a_{0}da_{1}...da_{n}:=a_{0}\left[ F,a_{1}\right] ...\left[ F,a_{n}\right] ~.
\]

It must be emphasized that the noncommutative generalization of differential
calculus is a wide and wild generalization since it enables us to extend
differential calculus to fractals!

\section{NC Riemannian geometry, Clifford algebras, and Dirac operator}

Another great achievement of Alain Connes was the complete and deep
reinterpretation of the $ds^{2}$ in Riemannian geometry. Classically, $%
ds^{2}=g_{\mu \nu }dx^{\mu }dx^{\nu }$.\ In the noncommutative framework, $dx
$ must be interpreted as $dx=[F,x]$ (where $(\mathcal{A},\mathcal{H},F)$ is
a Fredholm module), and the matrix $\left( g_{\mu \nu }\right) $ as an
element of the $n\times n$ matrix algebra $M_{n}(\mathcal{A})$. The $ds^{2}$
must therefore become a \emph{compact} and \emph{positive} operator of the
form 
\[
G=[F,x^{\mu }]^{\ast }g_{\mu \nu }[F,x^{\nu }]~.
\]

\subsection{A redefinition of distance}

Connes' idea is to reinterpret the classical definition of distance $d(p,q)$
between two points $p,q$ of a Riemannian manifold $M$ as the $\func{Inf}$ of
the length $L(\gamma )$ of the paths $\gamma :p\rightarrow q$%
\[
d(p,q)=\underset{\gamma :p\rightarrow q}{\func{Inf}}L(\gamma ) 
\]

\[
L(\gamma )=\int_{p}^{q}ds=\int_{p}^{q}\left( g_{\mu \nu }dx^{\mu }dx^{\nu
}\right) ^{1/2}~.
\]

\noindent Using the equivalence between a point $x$ of $M$ and the pure
state $\delta _{x}$ on the commutative $C^{*}$-algebra $\mathcal{A}%
:=C^{\infty }\left( M\right) $, an elementary computation shows that this
definition of the distance is equivalent to the dual algebraic definition
using only concepts concerning the $C^{*}$-algebra $\mathcal{A}$%
\[
d(p,q)=\func{Sup}\left\{ \left| f(q)-f(p)\right| :\left\| \func{grad}%
(f)\right\| _{\infty }\leq 1\right\} 
\]

\noindent where $\left\Vert ...\right\Vert _{\infty }$ is the $L^{\infty }$
norm, that is the $\func{Sup}$ on $x\in M$ of the norms on the tangent
spaces $T_{x}M$.\footnote{%
Let $\gamma :I=\left[ 0,1\right] \rightarrow M$ be a $C^{\infty }$ curve in $%
M$ from $p$ to $q$. $L(\gamma )=\int_{p}^{q}\left\vert \dot{\gamma}\left(
t\right) \right\vert dt=\int_{0}^{1}g\left( \dot{\gamma}\left( t\right) ,%
\dot{\gamma}\left( t\right) \right) ^{1/2}dt$. If $f\in C^{\infty }\left(
M\right) $, using the duality between $df$ and $\func{grad}f$ induced by the
metric, we find $f(q)-f(p)=\int_{0}^{1}df_{\gamma (t)}\left( \dot{\gamma}%
\left( t\right) \right) dt=\int_{0}^{1}g_{\gamma (t)}\left( \func{grad}%
_{\gamma (t)}f,\dot{\gamma}\left( t\right) \right) dt$. This shows that $%
\left\vert f(q)-f(p)\right\vert \leq \int_{0}^{1}\left\vert \func{grad}%
_{\gamma (t)}f\right\vert \left\vert \dot{\gamma}\left( t\right) \right\vert
dt\leq \left\Vert \func{grad}f\right\Vert _{\infty }L\left( \gamma \right) $%
. Therefore, if $\left\Vert \func{grad}(f)\right\Vert _{\infty }\leq 1$ we
have $\left\vert f(q)-f(p)\right\vert \leq d(p,q).\;$When we take the Sup we
retreive $d(p,q)$ using the special function $f_{p}(x)=d(p,x)$ since $%
\left\vert f_{p}(q)-f_{p}(p)\right\vert =d(p,q)$.}

\subsection{Clifford algebras}

Now the core of the noncommutative definition of distance is the use of the 
\emph{Dirac operator}.\ In order to explain this key point, which transforms
the geometrical concept of distance into a quantum concept, the \emph{%
Clifford algebra} of a Riemannian manifold must be introduced.

Recall that the formalism of Clifford algebras relates the differential
forms and the metric on a Riemannian manifold. In the simple case of
Euclidean space $\mathbb{R}^{n}$, the main idea is to encode the isometries $%
O(n)$ in an algebra structure.\ Since every isometry is a product of
reflections (Cartan), one can associate to any vector $v\in \mathbb{R}^{n}$
the reflection $\overline{v}$ relative to the orthogonal hyperplane $v^{\bot
}$ and introduce a multiplication $v.w$ which is nothing else than the
composition $\overline{v}\circ \overline{w}$. We are then naturally led to
the anti-commutation relations 
\[
\left\{ v,w\right\} :=v.w+w.v=-2(v,w)
\]

\noindent where $(v,w)$ is the Euclidean scalar product.

More generally, let $V$ be a $\mathbb{R}$-vector space endowed with a
quadratic form $g$.\ Its Clifford algebra $\func{Cl}(V,g)$ is its tensor
algebra $\mathcal{T}(V)=\oplus _{k=0}^{k=\infty }V^{\otimes k}$ quotiented
by the relations 
\[
v\otimes v=-g(v)1,\;\forall v\in V 
\]

\noindent (where $g(v)=g(v,v)=\left\| v\right\| ^{2}$). In $\func{Cl}(V,g)$
the tensorial product $v\otimes v$ becomes a product $v.v=v^{2}$. It must be
stressed that there exists always in $\func{Cl}(V,g)$ the constants $\mathbb{%
R}$ which correspond to the $0$th tensorial power of $V$.

Using the scalar product 
\[
2g(v,w)=g(v+w)-g(v)-g(w) 
\]

\noindent one gets the anti-commutation relations 
\[
\left\{ v,w\right\} =-2g(v,w)~.
\]

Elementary examples are given by the $\func{Cl}_{n}=\func{Cl}(\mathbb{R}%
^{n},g_{\func{Euclid}}).$

\begin{itemize}
\item $\func{Cl}_{0}=\mathbb{R}$.

\item $\func{Cl}_{1}=\mathbb{C}$ ($V=i\mathbb{R}$, $i^{2}=-1$, $\func{Cl}%
_{1}=\mathbb{R\oplus }i\mathbb{R}$).

\item $\func{Cl}_{2}=\mathbb{H}$ ($V=i\mathbb{R+}j\mathbb{R}$, $ij=k$, $%
\func{Cl}_{2}=\mathbb{R\oplus }i\mathbb{R\oplus }j\mathbb{R\oplus }k\mathbb{R%
}$).

\item $\func{Cl}_{3}=\mathbb{H\oplus H}$.

\item $\func{Cl}_{4}=\mathbb{H}[2]$ ($2\times 2$ matrices with entries in $%
\mathbb{H}$).

\item $\func{Cl}_{5}=\mathbb{C}[4]$.

\item $\func{Cl}_{6}=\mathbb{R}[8]$.

\item $\func{Cl}_{7}=\mathbb{R}[8]\oplus \mathbb{R}[8]$.

\item $\func{Cl}_{n+8}=\func{Cl}_{n}\otimes \mathbb{R}[16]$ (Bott
periodicity theorem).
\end{itemize}

If $g(v)\neq 0$ (which would always be the case for $v\neq 0$ if $g$ is non
degenerate) $v$ is invertible in this algebra structure and 
\[
v^{-1}=-\frac{v}{g(v)}~.
\]

\noindent The multiplicative Lie group $\func{Cl}^{\times }(V,g)$ of the
invertible elements of $\func{Cl}(V,g)$ act through inner automorphisms on $%
\func{Cl}(V,g)$.\ This yields the adjoint representation 
\[
\begin{array}{rll}
\func{Ad}:\func{Cl}^{\times }(V,g) & \rightarrow  & \func{Aut}\left( \func{Cl%
}(V,g)\right)  \\ 
v & \mapsto  & \func{Ad}_{v}:w\mapsto v.w.v^{-1}~%
\end{array}%
\]

\noindent But~\footnote{$v.w.v^{-1}=$ $-v.w.\frac{v}{g(v)}=$ $-(-w.v-2g(v,w))%
\frac{v}{g(v)}=$ $w.\frac{v^{2}}{g(v)}+\frac{2g(v,w)v}{g(v)}=$ $-w+\frac{%
2g(v,w)v}{g(v)}.$} 
\[
v.w.v^{-1}=-w+\frac{2g(v,w)v}{g(v)}=\func{Ad}_{v}(w)~.
\]

\noindent As $-\func{Ad}_{v}$ is the reflection relative to $v^{\bot }$,
this means that reflections act through the adjoint representation of the
Clifford algebra. The derivative $\func{ad}$ of the adjoint representation
enables to recover the Lie bracket of the Lie algebra $\func{cl}^{\times
}(V,g)=\func{Cl}(V,g)$ of the Lie group $\func{Cl}^{\times }(V,g)$%
\[
\begin{array}{rll}
\func{ad}:\func{cl}^{\times }(V,g)=\func{Cl}(V,g) & \rightarrow & \func{Der}%
\left( \func{Cl}(V,g)\right) \\ 
v & \mapsto & \func{ad}_{v}:w\mapsto [v,w]%
\end{array}
\]

Now there exists a fundamental relation between the Clifford algebra $(V,g)$
of $V$ and its exterior algebra $\Lambda ^{\ast }V$.\ If $g=0$ and if we
interpret $v.w$ as $v\wedge w$, the anti-commutation relations become simply 
$\{v,w\}=0$, which is the classical antisymmetry $w\wedge v=-v\wedge w$ of
differential $1$-forms. Therefore 
\[
\Lambda ^{\ast }V=(V,0)~.
\]

\noindent In fact, $(V,g)$ can be considered as a way of \emph{quantizing} $%
\Lambda ^{\ast }V$ using the metric $g$ in order to get non trivial
anti-commutation relations.

Due to the relations $v^{2}=-g(v)1$ which decrease the degree of a product
by $2$, $\func{Cl}(V,g)$ is no longer a $\mathbb{Z}$-graded algebra but only
a $\mathbb{Z}/2$-graded algebra, the $\mathbb{Z}/2$-gradation corresponding
to the even/odd elements. But we can reconstruct a $\mathbb{Z}$-graded
algebra $\mathcal{C}=\bigoplus\limits_{k=0}^{k=\infty }C^{k}$ associated to $%
\func{Cl}(V,g)$, the $C^{k}$ being the homogeneous terms of degree $k$: $%
v_{1}.\cdots .v_{k}$.

\textbf{Theorem}.\ The map of graded algebras $\mathcal{C}%
=\bigoplus\limits_{k=0}^{k=\infty }C^{k}\rightarrow \Lambda ^{\ast
}V=\bigoplus\limits_{k=0}^{k=\infty }\Lambda ^{k}$ given by $v_{1}.\cdots
.v_{k}\rightarrow v_{1}\wedge \cdots \wedge v_{k}$ is a \emph{linear}
isomorphism (but not an \emph{algebra} isomorphism).

We consider now $2$ operations on the exterior algebra $\Lambda ^{\ast }V$ :

\begin{enumerate}
\item The outer multiplication $\varepsilon (v)$ by $v\in V$: 
\[
\varepsilon (v)\left( \underset{i}{\wedge }u_{i}\right) =v\wedge \left( 
\underset{i}{\wedge }u_{i}\right) ~.\ 
\]%
We have $\varepsilon (v)^{2}=0$ since $v\wedge v=0$.

\item The contraction (inner multiplication) $\iota (v)$ induced by the
metric $g$:\footnote{%
In the following formula $\widehat{u_{j}}$ means that the term $u_{j}$ is
deleted.} 
\[
\iota (v)\left( \underset{i}{\wedge }u_{i}\right)
=\sum\limits_{j=1}^{j=k}(-1)^{j}g(v,u_{j})\;u_{1}\wedge \cdots \wedge 
\widehat{u_{j}}\wedge \cdots u_{k}~.
\]

We have also $\iota (v)^{2}=0$. The inner multiplication $\iota (v)$ is a
supplementary structure involving the metric structure.
\end{enumerate}

\noindent One shows that the following anti-commutations relations obtain: 
\[
\left\{ \varepsilon (v),\iota (w)\right\} =-g(v,w)1~.
\]

\noindent Let now $c(v)=\varepsilon (v)+\iota (v)$.\ We get the
anti-commutation relations of the Clifford algebra 
\[
\left\{ c(v),c(w)\right\} =-2g(v,w)1
\]

\noindent and $\func{Cl}(V,g)$ is therefore generated in $\func{End}_{%
\mathbb{R}}\left( \Lambda ^{*}V\right) $ by the $c(v)$ (identified with $v$).

\subsection{Spin groups}

The isometry group $O(n)$ is canonically embedded in $\func{Cl}(V,g)$ since
every isometry is a product of reflections. In fact $\func{Cl}(V,g)$
contains also the \emph{pin group} $\func{Pin}(n)$ which is a $2$-fold
covering of $O(n)$.\ If we take into account the orientation and restrict to 
$SO(n)$, the $2$-fold covering becomes the \emph{spin group} $\func{Spin}(n)$%
. $\func{Spin}(n)$ is generated by the even products of $v$ s.t. $g(v)=\pm 1$%
, $SO(n)$ is generated by even products of $-Ad_{v}$ and the covering $\func{%
Spin}(n)\rightarrow SO(n)$ is given by $v\mapsto -Ad_{v}$. By restriction of
the Clifford multiplication and of the adjoint representation $w\mapsto
v.w.v^{-1}$ to $\func{Spin}(n)$, we get therefore a representation $\gamma $
of $\func{Spin}(n)$ into the spinor space $\mathbb{S}=\func{Cl}(V,g)$.

\subsection{Dirac equation}

We can use the Clifford algebra, and therefore the metric, to change the
classical exterior derivative of differential forms given by 
\[
d:=\varepsilon \left( dx^{\mu }\right) \frac{\partial }{\partial x^{\mu }}~.
\]

\noindent We then define the Dirac operator on spinor fields $\mathbb{R}%
^{n}\rightarrow \mathbb{S}$ as 
\begin{eqnarray*}
D &:&=c\left( dx^{\mu }\right) \frac{\partial }{\partial x^{\mu }} \\
&=&\gamma ^{\mu }\frac{\partial }{\partial x^{\mu }}~.
\end{eqnarray*}

\noindent where $c$ is the Clifford multiplication,\ and $D$ acts on the
spinor space $\mathbb{S}=\func{Cl}(V,g)$. As $\left\{ c(v),c(w)\right\}
=-2g(v,w)1$, the $\gamma ^{\mu }$ satisfy standard Dirac relations of
anticommutation $\left\{ \gamma ^{\mu },\gamma ^{\nu }\right\} =-2\delta
^{\mu \nu }$ in the Euclidean case.\footnote{%
The classical Dirac matrices are the $-i\gamma ^{\mu }$ for $\mu =0,1,2,3$.}
On can check that $D^{2}=\Delta $ is the Laplacian.

\subsection{Dirac operator}

More generally, if $M$ is a Riemannian manifold, the previous construction
can be done for every tangent space $T_{x}M$ endowed with the quadratic form 
$g_{x}$.\ In this way we get a bundle of Clifford algebras $\func{Cl}(TM,g)$%
. If $S$ is a spinor bundle, that is a bundle of $\func{Cl}(TM)$ -modules
s.t. $\func{Cl}(TM)\simeq \func{End}(S)$, endowed with a covariant
derivative $\nabla $, we associate to it the Dirac operator 
\[
D:\mathcal{S}=\Gamma (S)=C^{\infty }(M,S)\rightarrow \Gamma (S)
\]

\noindent which is a first order elliptic operator interpretable as the
``square root'' of the Laplacian $\Delta $, $\Delta $ interpreting itself
the metric in operatorial terms. The Dirac operator $D$ establishes a
coupling between the covariant derivation on $S$ and the Clifford
multiplication of $1$-forms. It can be extended from the $C^{\infty }(M)$%
-module $\mathcal{S}=\Gamma (S)$ to the Hilbert space $\mathcal{H}
=L^{2}(M,S) $.

In general, due to chirality, $S$ will be the direct sum of an even and an
odd part, $S=S^{+}\oplus S^{-}$ and $D$ will have the characteristic form 
\begin{eqnarray*}
D &=&\left[ 
\begin{array}{cc}
0 & D^{-} \\ 
D^{+} & 0%
\end{array}%
\right]  \\
D^{+} &:&\Gamma (S^{+})\rightarrow \Gamma (S^{+}) \\
D^{-} &:&\Gamma (S^{-})\rightarrow \Gamma (S^{-})
\end{eqnarray*}

\noindent $D^{+}$ and $D^{-}$ being adjoint operators.

\subsection{Noncommutative distance and Dirac operator}

In this classical framework, it easy to compute the bracket $[D,f]$ for $%
f\in C^{\infty }(M)$.\ First, there exists on $M$ the \emph{Levi-Civita
connection}: 
\[
\nabla ^{g}:\Omega ^{1}(M)\rightarrow \Omega ^{1}(M)\underset{C^{\infty }(M)}%
{\otimes }\Omega ^{1}(M)
\]

\noindent satisfying the Leibniz rule for $\alpha \in \Omega ^{1}(M)$ and $%
f\in C^{\infty }(M)$: 
\[
\nabla ^{g}(\alpha f)=\nabla ^{g}(\alpha )f+\alpha \otimes df 
\]

\noindent (as $\nabla ^{g}(\alpha )\in \Omega ^{1}(M)\underset{C^{\infty }(M)%
}{\otimes }\Omega ^{1}(M)$, $\nabla ^{g}(\alpha )f\in \Omega ^{1}(M)\underset%
{C^{\infty }(M)}{\otimes }\Omega ^{1}(M)$ and as $\alpha $ and $df\in \Omega
^{1}(M)$, $\alpha \otimes df\in \Omega ^{1}(M)\underset{C^{\infty }(M)}{%
\otimes }\Omega ^{1}(M)$). There exists also the \emph{spin connection} on
the spinor bundle $S$ 
\[
\nabla ^{S}:\Gamma (S)\rightarrow \Omega ^{1}(M)\underset{C^{\infty }(M)}{%
\otimes }\Gamma (S) 
\]

\noindent satisfying the Leibniz rule for $\psi \in \Gamma (S)$ and $f\in
C^{\infty }(M)$: 
\begin{eqnarray*}
\nabla ^{S}(\psi f) &=&\nabla ^{S}(\psi )f+\psi \otimes df \\
\nabla ^{S}\left( \gamma (\alpha )\psi \right) &=&\gamma \left( \nabla
^{g}(\alpha )\right) \psi +\gamma (\alpha )\nabla ^{S}(\psi )
\end{eqnarray*}

\noindent where $\gamma $ is the spin representation. 

The Dirac operator on $\mathcal{H}=L^{2}(M,S)$ is then defined as 
\[
D:=\gamma \circ \nabla ^{S}~.
\]

\noindent If $\psi \in \Gamma (S)$, we have (making the $f$ acting on the
left in $\mathcal{H}$) 
\begin{eqnarray*}
D\left( f\psi \right)  &=&\gamma \left( \nabla ^{S}(\psi f)\right)  \\
&=&\gamma \left( \nabla ^{S}(\psi )f+\psi \otimes df\right)  \\
&=&\gamma \left( \nabla ^{S}(\psi )\right) f+\gamma \left( \psi \otimes
df\right)  \\
&=&fD(\psi )+\gamma \left( df\right) \psi 
\end{eqnarray*}

\noindent and therefore $[D,f](\psi )=fD(\psi )+\gamma \left( df\right) \psi
-fD(\psi )=\gamma \left( df\right) \psi $, that is 
\[
\lbrack D,f]=\gamma \left( df\right) ~.
\]

In the standard case where $M=\mathbb{R}^{n}$ and $S=\mathbb{R}^{n}\times V$%
, $V$ being a $\func{Cl}_{n}$-module of spinors ($\func{Cl}_{n}=\func{Cl}%
\left( \mathbb{R}^{n},g_{\func{Euclid}}\right) $), we have seen that $D$ is
a differential operator with constant coefficients taking its values in $V$. 
\[
D=\sum\limits_{k=1}^{k=n}\gamma ^{\mu }\frac{\partial }{\partial x^{\mu }} 
\]

\noindent with the constant matrices $\gamma ^{\mu }\in \mathcal{L}(V)$
satisfying the anti-commutation relations 
\[
\left\{ \gamma ^{\mu },\gamma ^{\nu }\right\} =-2\delta ^{\mu \nu }~.
\]

\noindent The fundamental point is that the $\gamma ^{\mu }$ are associated
with the basic $1$-forms $dx^{\mu }$ through the isomorphism 
\[
c:\mathcal{C}=\Lambda ^{*}(M)\rightarrow \func{gr}\left( \func{Cl}
(TM)\right) 
\]

\[
\lbrack D,f]=\gamma \left( df\right) =c(df) 
\]

\noindent and $\left\Vert [D,f]\right\Vert $ is the norm of the Clifford
action of $df$ on the space of spinors $L^{2}(M,S)$. But 
\begin{eqnarray*}
\left\Vert c(df)\right\Vert ^{2} &=&\underset{x\in M}{\func{Sup}}%
g_{x}^{-1}\left( d\overline{f}(x),df(x)\right)  \\
&=&\underset{x\in M}{\func{Sup}}g_{x}\left( \limfunc{grad}\nolimits_{x}%
\overline{f},\limfunc{grad}\nolimits_{x}f\right)  \\
&=&\left\Vert \func{grad}(f)\right\Vert _{\infty }^{2}~.
\end{eqnarray*}

\noindent Whence the definition: 
\[
d(p,q)=\func{Sup}\left\{ \left\vert f(p)-f(q)\right\vert :f\in \mathcal{A}%
,\left\Vert [D,f]\right\Vert \leq 1\right\} ~.
\]

In this reinterpretation, $ds$ corresponds to \emph{the propagator of the
Dirac operator} $D.$ As an operator acting on the Hilbert space $\mathcal{H}$%
, $D$ is an unbounded self-adjoint operator such that $[D,f]$ is bounded for
every $f\in \mathcal{A}$ and such that its resolvent $(D-\lambda I)^{-1}$ is
compact for every $\lambda \notin \func{Sp}(D)$ (which corresponds to the
fact that $ds$ is infinitesimal) and the trace $\func{Trace}\left(
e^{-D^{2}}\right) $ is \emph{finite}. In terms of the operator $G=[F,x^{\mu
}]^{\ast }g_{\mu \nu }[F,x^{\nu }]$, we have $G=D^{-2}$.

\section{Noncommutative spectral geometry}

Basing himself on several examples, Alain Connes arrived at the following
concept of noncommutative geometry. In the classical commutative case, $%
\mathcal{A}=C^{\infty }\left( M\right) $ is the commutative algebra of
\textquotedblleft coordinates\textquotedblright\ on $M$ represented in the
Hilbert space $\mathcal{H}=L^{2}(M,S)$ by pointwise multiplication~\footnote{%
If $f\in \mathcal{A}$ and $\xi \in \mathcal{H}$, $\left( f\xi \right)
(x)=f(x)\xi (x)$.} and $ds$ is a symbol non commuting with the $f\in 
\mathcal{A}$ and satisfying the commutation relations $\left[ \left[
f,ds^{-1}\right] ,g\right] =0$, for every $f,g\in \mathcal{A}$. Any specific
geometry is defined through the representation $ds=D^{-1}$ of $ds$ by means
of a Dirac operator $D=\gamma ^{\mu }\nabla _{\mu }$. The differential $df=%
\left[ D,f\right] $ is then the Clifford multiplication by the gradient $%
\nabla f$ and its norm in $\mathcal{H}$ is the Lipschitz norm of $f$: $%
\left\Vert \left[ D,f\right] \right\Vert =\underset{x\in M}{\func{Sup}}%
\left\Vert \nabla f\right\Vert $.

These results can be taken as a definition in the general case.\ The
geometry is defined by a spectral triple $\left( \mathcal{A},\mathcal{H}%
,D\right) $ where $\mathcal{A}$ is a noncommutative$\;C^{\ast }$-algebra
with a representation in an Hilbert space $\mathcal{H}$ and $D$ is an
unbounded self-adjoint operator on $\mathcal{H}$ such that $ds=D^{-1}$ and
more generally the resolvent $\left( D-\lambda I\right) ^{-1}$, $\lambda
\notin \mathbb{R}$, is compact, and at the same time all $\left[ D,a\right] $
are bounded for every $a\in \mathcal{A}$ (there is a tension between these
two last conditions). As Connes \cite{Connes2000b} emphasizes

\begin{quotation}
\noindent 
``It is precisely this lack of commutativity between the line element and
the coordinates on a space [between $ds$ and the $a\in \mathcal{A}$] that
will provide the measurement of distance.''
\end{quotation}

\noindent The new definition of differentials are then $da=\left[ D,a\right] 
$ for any $a\in \mathcal{A}$.

\section{Yang-Mills theory of a NC coupling between $2$ points and Higgs
mechanism}

A striking example of pure noncommutative physics is given by Connes'
interpretation of the Higgs phenomenon.\ In the Standard Model, the Higgs
mechanism was an \emph{ad hoc} device used for confering a mass to gauge
bosons.\ It lacked any geometrical interpretation.\ One of the deepest
achievement of the noncommutative perspective has been to show that Higgs
fields correspond effectively to gauge bosons for a \emph{discrete}
noncommutative geometry.

\subsection{Symmetry breaking and classical Higgs mechanism}

Let us first recall the classical Higgs mechanism.\ Consider e.g. a $\varphi
^{4}$ theory for $2$ scalar real fields $\varphi _{1}$ and $\varphi _{2}$.
The Lagrangian is 
\[
\mathcal{L}=\frac{1}{2}\left( \partial _{\mu }\varphi _{1}\partial ^{\mu
}\varphi _{1}+\partial _{\mu }\varphi _{2}\partial ^{\mu }\varphi
_{2}\right) -V\left( \varphi _{1}^{2}+\varphi _{2}^{2}\right) 
\]

\noindent with the quartic potential 
\[
V\left( \varphi _{1}^{2}+\varphi _{2}^{2}\right) =\frac{1}{2}\mu ^{2}\left(
\varphi _{1}^{2}+\varphi _{2}^{2}\right) +\frac{1}{4}\left\vert \lambda
\right\vert \left( \varphi _{1}^{2}+\varphi _{2}^{2}\right) ^{2}~.
\]

\noindent It is by construction $SO(2)$-invariant.

For $\mu ^{2}>0$ the minimum of $V$ (the quantum vacuum) is non degenerate: $%
\varphi _{0}=(0,0)$ and the Lagrangian $\mathcal{L}_{os}$ of small
oscillations in the neighborhood of $\varphi _{0}$ is the sum of $2$
Lagrangians of the form: 
\[
\mathcal{L}_{os}=\frac{1}{2}\left( \partial _{\mu }\psi \partial ^{\mu }\psi
\right) -\frac{1}{2}\mu ^{2}\psi ^{2} 
\]

\noindent describing particles of mass $\mu ^{2}$.

But for $\mu ^{2}<0$ the situation becomes completely different. Indeed the
potential $V$ has a full circle (a $SO(2)$-orbit) of minima 
\[
\varphi _{0}^{2}=-\frac{\mu ^{2}}{\left\vert \lambda \right\vert }=v^{2}
\]

\noindent and the vacuum state is highly \emph{degenerate}.

One must therefore \emph{break the symmetry} to choose a vacuum state.\ Let
us take for instance $\varphi _{0}=\left[ 
\begin{array}{l}
v \\ 
0%
\end{array}%
\right] $ and translate the situation to $\varphi _{0}$: 
\[
\varphi =\left[ 
\begin{array}{l}
\varphi _{1} \\ 
\varphi _{2}%
\end{array}%
\right] =\left[ 
\begin{array}{l}
v \\ 
0%
\end{array}%
\right] +\left[ 
\begin{array}{l}
\xi  \\ 
\eta 
\end{array}%
\right] ~.
\]

\noindent The oscillation Lagrangian at $\varphi _{0}$ becomes 
\[
\mathcal{L}_{os}=\frac{1}{2}\left( \partial _{\mu }\eta \partial ^{\mu }\eta
+2\mu ^{2}\eta ^{2}\right) +\frac{1}{2}\left( \partial _{\mu }\xi \partial
^{\mu }\xi \right) 
\]

\noindent and describes $2$ particles:

\begin{enumerate}
\item a particle $\eta $ of mass $m=\sqrt{2}\left| \mu \right| $, which
corresponds to radial oscillations,

\item a particule $\xi $ of mass $m=0$, which connects vacuum states. $\xi $
is the \emph{Goldstone boson}.
\end{enumerate}

As is well known, the Higgs mechanism consists in using a cooperation
between gauge bosons and Goldstone bosons to confer a mass to gauge bosons.
Let $\varphi =\frac{1}{\sqrt{2}}\left( \varphi _{1}+i\varphi _{2}\right) $
be the scalar complex field associated to $\varphi _{1}$ and $\varphi _{2}$%
.\ Its Lagrangian is 
\[
\mathcal{L}=\partial _{\mu }\overline{\varphi }\partial ^{\mu }\varphi -\mu
^{2}\left\vert \varphi \right\vert ^{2}-\left\vert \lambda \right\vert
\left\vert \varphi \right\vert ^{4}~.
\]

\noindent It is trivially invariant by the global internal symmetry $\varphi
\rightarrow e^{i\theta }\varphi $. If we \emph{localize} the global symmetry
using transformations $\varphi (x)\rightarrow e^{iq\alpha (x)}\varphi (x)$
and take into account the coupling with an electro-magnetic field deriving
from the vector potential $A_{\mu }$, we get 
\[
\mathcal{L}=\nabla _{\mu }\overline{\varphi }\nabla ^{\mu }\varphi -\mu
^{2}\left| \varphi \right| ^{2}-\left| \lambda \right| \left| \varphi
\right| ^{4}-\frac{1}{4}F_{\mu \nu }F^{\mu \nu } 
\]

\noindent where $\nabla $ is the covariant derivative 
\[
\nabla _{\mu }=\partial _{\mu }+iqA_{\mu } 
\]

\noindent and $F$ the force field 
\[
F_{\mu \nu }=\partial _{\nu }A_{\mu }-\partial _{\mu }A_{\nu }~.
\]

\noindent The Lagrangian remains invariant if we balance the localization of
the global internal symmetry with a change of gauge 
\[
A_{\mu }\rightarrow A_{\mu }^{\prime }=A_{\mu }-\partial _{\mu }\alpha (x)~.
\]

For $\mu ^{2}>0$, $\varphi _{0}=0$ is a minimum of $V(\varphi ),$ the vacuum
is non degenerate, and we get $2$ scalar particles $\varphi $ and $\overline{%
\varphi }$ and a photon $A_{\mu }$.

For $\mu ^{2}<0$, the vacuum is degenerate and there is a spontaneous
symmetry breaking.\ We have $\left| \varphi _{0}\right| ^{2}=-\frac{\mu ^{2}%
}{2\left| \lambda \right| }=\frac{v^{2}}{2}$.$\;$If we take $\varphi _{0}=%
\frac{^{v}}{\sqrt{2}}$ and write 
\[
\varphi =\varphi ^{\prime }+\varphi _{0}=\frac{1}{\sqrt{2}}(v+\eta +i\xi
)\approx \frac{1}{\sqrt{2}}e^{i\frac{\xi }{v}}(v+\eta )\text{ for }\xi \text{
and }\eta \text{ small,} 
\]

\noindent we get for the Lagrangian of oscillations: 
\[
\mathcal{L}_{os}=\frac{1}{2}\left( \partial _{\mu }\eta \partial ^{\mu }\eta
+2\mu ^{2}\eta ^{2}\right) +\frac{1}{2}\left( \partial _{\mu }\xi \partial
^{\mu }\xi \right) -\frac{1}{4}F_{\mu \nu }F^{\mu \nu }+qvA_{\mu }\left(
\partial _{\mu }\xi \right) +\frac{q^{2}v^{2}}{2}A_{\mu }A^{\mu }~.
\]

\begin{enumerate}
\item The field $\eta $ (radial oscillations) has mass $m=\sqrt{2}\left| \mu
\right| .$

\item The boson $A_{\mu }$ acquires a mass due to the term $A_{\mu }A^{\mu } 
$ and interacts with the Goldstone boson $\xi $.
\end{enumerate}

The terms containing the gauge boson $A_{\mu }$ and the Goldstone boson $\xi 
$ write 
\[
\frac{q^{2}v^{2}}{2}\left( A_{\mu }+\frac{1}{qv}\partial _{\mu }\xi \right)
\left( A^{\mu }+\frac{1}{qv}\partial ^{\mu }\xi \right) 
\]

\noindent and are therefore generated by the gauge change 
\begin{eqnarray*}
\alpha  &=&\frac{\xi }{qv} \\
A_{\mu } &\rightarrow &A_{\mu }+\partial _{\mu }\alpha ~.
\end{eqnarray*}

We see that we can use the gauge transformations 
\[
A_{\mu }\rightarrow A_{\mu }^{\prime }=A_{\mu }+\frac{1}{qv}\partial ^{\mu
}\xi 
\]%
for \emph{fixing} the vacuum state.\ The transformation corresponds to the
phase rotation of the scalar field 
\[
\varphi \rightarrow \varphi ^{\prime }=e^{-i\frac{\xi }{v}}\varphi =\frac{%
v+\eta }{\sqrt{2}}~.
\]

In this new gauge where the Goldstone boson $\xi $ disappears, the vector
particule $A_{\mu }^{\prime }$ acquires a mass $qv$. The Lagrangian writes
now 
\[
\mathcal{L}_{os}=\frac{1}{2}\left( \partial _{\mu }\eta \partial ^{\mu }\eta
+2\mu ^{2}\eta ^{2}\right) -\frac{1}{4}F_{\mu \nu }F^{\mu \nu }+\frac{%
q^{2}v^{2}}{2}A_{\mu }^{\prime }A^{\prime \mu }~.
\]

\noindent The Goldstone boson connecting the degenerate vacuum states is in
some sense ``captured'' by the gauge boson and transformed into mass.

\subsection{Noncommutative Yang-Mills theory of $2$ points and Higgs
phenomenon}

The noncommutative equivalent of this description is the following.\ It
shows that Higgs mechanism is actually the standard Yang-Mills formalism
applied to a \emph{purely discrete noncommutative geometry}.

Let $\mathcal{A}=\mathcal{C}(Y)=\mathbb{C}\oplus \mathbb{C}$ be the $C^{*}$%
-algebra of the space $Y$ composed of $2$ points $a$ and $b$. Its elements $%
f=\left[ 
\begin{array}{cc}
f(a) & 0 \\ 
0 & f(b)%
\end{array}
\right] $ act through multiplication on the Hilbert space $\mathcal{H}=%
\mathcal{H}_{a}\oplus \mathcal{H}_{b}$. We take for Dirac operator an
operator of the form 
\[
D=\left[ 
\begin{array}{cc}
0 & M^{*}=D^{-} \\ 
M=D^{+} & 0%
\end{array}
\right] 
\]

\noindent and introduce the ``chirality'' $\gamma =\left[ 
\begin{array}{cc}
1 & 0 \\ 
0 & -1%
\end{array}
\right] $ (the $\gamma _{5}$ of the standard Dirac theory). In this discrete
situation we define $df$ as 
\[
df=[D,f]=\Delta f\left[ 
\begin{array}{cc}
0 & M^{*} \\ 
-M & 0%
\end{array}
\right] 
\]

\noindent with $\Delta f=f(b)-f(a).\;$Therefore 
\[
\left\| \lbrack D,f]\right\| =\left| \Delta f\right| \lambda 
\]
where $\lambda =\left\| M\right\| $ is the greatest eigenvalue of $M$.

If we apply now the formula for the distance, we find: 
\begin{eqnarray*}
d(a,b) &=&\func{Sup}\left\{ \left| f(a)-f(b)\right| :f\in \mathcal{A}%
,\left\| [D,f]\right\| \leq 1\right\} \\
&=&\func{Sup}\left\{ \left| f(a)-f(b)\right| :f\in \mathcal{A},\left|
f(a)-f(b)\right| \lambda \leq 1\right\} \\
&=&\frac{1}{\lambda }
\end{eqnarray*}

\noindent and we see that the distance $\frac{1}{\lambda }$ between the two
points $a$ and $b$ has a \emph{spectral} content and is measured by the
Dirac operator.

To interpret differential calculus in this context, we consider the $2$
idempotents (projectors) $e$ and $1-e$ defined by 
\begin{eqnarray*}
e(a) &=&1,e(b)=0 \\
(1-e)(a) &=&0,(1-e)(b)=1~.
\end{eqnarray*}

\noindent Every $f\in \mathcal{A}$ writes $f=f(a)e+f(b)(1-e)$, and therefore 
\begin{eqnarray*}
df &=&f(a)de+f(b)d(1-e) \\
&=&\left( f(a)-f(b)\right) de \\
&=&-(\Delta f)de \\
&=&-(\Delta f)ede+(\Delta f)(1-e)d(1-e)~.
\end{eqnarray*}

\noindent This shows that $ede$ and $(1-e)d(1-e)=-(1-e)de$ provide a natural
basis of the space of $1$-forms $\Omega ^{1}\mathcal{A}$.\ Let 
\begin{eqnarray*}
\omega &=&\lambda ede+\mu (1-e)d(1-e) \\
&=&\lambda ede-\mu (1-e)de
\end{eqnarray*}

\noindent a $1$-form.\ $\omega $ is represented by 
\[
\omega =\left( \lambda e-\mu (1-e)\right) [D,e]~.
\]

\noindent But on $\mathcal{H}$ $[D,e]=-\left[ 
\begin{array}{cc}
0 & M^{\ast } \\ 
-M & 0%
\end{array}%
\right] $ and therefore 
\[
\omega =\left[ 
\begin{array}{cc}
0 & -\lambda M^{\ast } \\ 
-\mu M & 0%
\end{array}%
\right] ~.
\]

Let us now construct with Connes the Yang-Mills theory corresponding to this
situation. A vector potential $V$~--- a connection in the sense of gauge
theories~--- is a self-adjoint $1$-form and has the form 
\begin{eqnarray*}
V &=&-\overline{\varphi }ede+\varphi (1-e)de \\
&=&\left[ 
\begin{array}{cc}
0 & \overline{\varphi }M^{\ast } \\ 
\varphi M & 0%
\end{array}%
\right] ~.
\end{eqnarray*}

\noindent Its curvature is the $2$-form 
\[
\theta =dV+V\wedge V 
\]

\noindent and an easy computation gives 
\[
\theta =-\left( \varphi +\overline{\varphi }+\varphi \overline{\varphi }%
\right) \left[ 
\begin{array}{cc}
-M^{\ast }M & 0 \\ 
0 & -MM^{\ast }%
\end{array}%
\right] ~.
\]

The Yang-Mills \emph{action} is the integral of the curvature $2$-form, that
is the trace of $\theta $:

\[
YM(V)=\func{Trace}\left( \theta ^{2}\right) ~.
\]

\noindent But as $\varphi +\overline{\varphi }+\varphi \overline{\varphi }
=\left| \varphi +1\right| ^{2}-1$ and 
\[
\func{Trace}\left( \left[ 
\begin{array}{cc}
-M^{*}M & 0 \\ 
0 & -MM^{*}%
\end{array}
\right] ^{2}\right) =2\func{Trace}\left( (M^{*}M)^{2}\right) 
\]

\noindent we get 
\[
YM(V)=2\left( \left\vert \varphi +1\right\vert ^{2}-1\right) ^{2}\func{Trace}%
\left( (M^{\ast }M)^{2}\right) ~.
\]

\subsection{Higgs mechanism}

This Yang-Mills action manifests a pure Higgs phenomenon of symmetry
breaking. The minimum of $YM(V)$ is reached everywhere on the circle $%
\left\vert \varphi +1\right\vert ^{2}=1$ (degeneracy) and the gauge group $%
\mathcal{U}=U(1)\times U(1)$ of the unitary elements of $\mathcal{A}$ acts
on it by 
\[
V\rightarrow uVu^{\ast }+udu^{\ast }
\]

\noindent where $u=\left[ 
\begin{array}{cc}
u_{1} & 0 \\ 
0 & u_{2}%
\end{array}
\right] $ with $u_{1},u_{2}\in U(1)$.

The field $\varphi $ is a Higgs bosonic field corresponding to a gauge
connection on a noncommutative space of $2$ points. If $\psi \in \mathcal{H}$
represents a fermionic state, the fermionic action is $I_{D}\left( V,\psi
\right) =\left\langle \psi ,\left( D+V\right) \psi \right\rangle $ with 
\[
D+V=\left[ 
\begin{array}{cc}
0 & \left( 1+\overline{\varphi }\right) M^{\ast } \\ 
\left( 1+\varphi \right) M & 0%
\end{array}%
\right] ~.
\]

\noindent The complete action coupling the fermion $\psi $ with the Higgs
boson $\varphi $ is therefore 
\[
YM(V)+I_{D}\left( V,\psi \right) ~.
\]

\section{The noncommutative derivation of the Glashow-Weinberg-Salam
Standard Model (Connes-Lott)}

A remarkable achievement of this noncommutative approach of Yang-Mills
theories is given by Connes-Lott's derivation of the Glashow-Weinberg-Salam
Standard Model. This derivation was possible because, as was emphasized by
Martin \emph{et al.} \cite{Martin} (p.\ 5), it ties

\begin{quotation}
\noindent 
``the properties of continuous spacetime with the intrinsic discreteness
stemming from the chiral structure of the Standard Model''.
\end{quotation}

\subsection{Gauge theory and NCG}

It is easy to reinterpret in the noncommutative framework classical gauge
theories where $M$ is a spin manifold, $\mathcal{A}=\mathcal{C}^{\infty }(M)$%
, $D$ is the Dirac operator and $\mathcal{H}=L^{2}(M,S)$ is the space of $%
L^{2}$ sections of the spinor bundle $S$. $\func{Diff}(M)=\func{Aut}(%
\mathcal{A})=\func{Aut}\left( \mathcal{C}^{\infty }(M)\right) $ is the
relativity group (the gauge group) of the theory: a diffeomorphism $\varphi
\in \func{Diff}(M)$ is identified with the $\ast $-automorphism $\alpha \in 
\func{Aut}(\mathcal{A})$ s.t. $\alpha \left( f\right) (x)=f\left( \varphi
^{-1}\left( x\right) \right) $. The main problem of quantum gravity is to
reconcile quantum field theory with general relativity, that is non abelian
gauge theories, which are noncommutative at the level of their \emph{internal%
} space of quantum variables, with the geometry of the \emph{external}
space-time $M$ with its group of diffeomorphism $\func{Diff}(M)$. The
noncommutative solution is an extraordinary principled one since it links
the standard \textquotedblleft inner\textquotedblright\ noncommutativity of
quantum internal degrees of freedom with the new \textquotedblleft
outer\textquotedblright\ noncommutativity of the external space.

\subsubsection{Inner automorphisms and internal symmetries}

The key fact is that, in the NC\ framework, there exists in $\func{Aut}(%
\mathcal{A})$ the normal subgroup $\func{Inn}(\mathcal{A})$ of \emph{inner
automorphisms} acting by conjugation $a\rightarrow uau^{-1}$.$\;\func{Inn}(%
\mathcal{A})$ is trivial in the commutative case and constitutes one of the
main feature of the NC case. As Alain Connes \cite{Connes96} emphasized:

\begin{quotation}
\noindent 
``Amazingly, in this description the group of gauge
transformation of the matter fields arises spontaneously as a normal
subgroup of the generalized diffeomorphism group $\func{Aut}(\mathcal{A})$.
It is the \emph{non commutativity} of the algebra $\mathcal{A}$ which gives
for free the group of gauge transformations of matter fields as a (normal)
subgroup of the group of diffeomorphisms.''
\end{quotation}

In $\func{Inn}(\mathcal{A})$ there exists in particular the \emph{unitary}
group $\mathcal{U}(\mathcal{A})$ of unitary elements $u^{*}=u^{-1}$ acting
by $\alpha _{u}\left( a\right) =uau^{*}$.

\subsubsection{Connections and vector potentials}

In the noncommutative framework we can easily reformulate standard
Yang-Mills theories. For that we need the concepts of a connection and of a
vector potential.

Let $\mathcal{E}$ be a finite projective (right) $\mathcal{A}$-module. A
connection $\nabla $ on $\mathcal{E}$ is a collection of morphisms (for
every $p)$ 
\[
\nabla :\mathcal{E}\otimes _{\mathcal{A}}\Omega ^{p}\left( \mathcal{A}%
\right) \rightarrow \mathcal{E}\otimes _{\mathcal{A}}\Omega ^{p+1}\left( 
\mathcal{A}\right) 
\]

\noindent satisfying for every $\omega \in \mathcal{E}\otimes _{\mathcal{A}
}\Omega ^{p}\left( \mathcal{A}\right) $ and every $\rho \in \Omega
^{q}\left( \mathcal{A}\right) $ the Leibniz rule in $\mathcal{E}\otimes _{%
\mathcal{A}}\Omega ^{p+q+1}\left( \mathcal{A}\right) $%
\[
\nabla \left( \omega \otimes \rho \right) =\nabla \left( \omega \right)
\otimes \rho +\left( -1\right) ^{p}\omega \otimes d\rho 
\]

\noindent where we use the relation $\Omega ^{p+1}\left( \mathcal{A}\right)
\otimes _{\mathcal{A}}\Omega ^{q}\left( \mathcal{A}\right) =\Omega
^{p}\left( \mathcal{A}\right) \otimes _{\mathcal{A}}\Omega ^{q+1}\left( 
\mathcal{A}\right) $. $\nabla $ is determined by its restriction to $\Omega
^{1}\left( \mathcal{A}\right) $

\[
\nabla :\mathcal{E}\otimes _{\mathcal{A}}\Omega ^{0}\left( \mathcal{A}%
\right) =\mathcal{E}\rightarrow \mathcal{E}\otimes _{\mathcal{A}}\Omega
^{1}\left( \mathcal{A}\right) 
\]

\noindent satisfying $\nabla \left( \xi a\right) =\nabla \left( \xi \right)
a+\xi \otimes da$ for $\xi \in \mathcal{E}$ and $a\in \mathcal{A}$.

The \emph{curvature} $\theta $ of $\nabla $ is given by $\nabla ^{2}:%
\mathcal{E}\rightarrow \mathcal{E}\otimes _{\mathcal{A}}\Omega ^{2}\left( 
\mathcal{A}\right) $.\ As

\begin{eqnarray*}
\nabla ^{2}\left( \xi a\right) &=&\nabla \left( \nabla \left( \xi \right)
a+\xi \otimes da\right) \\
&=&\nabla ^{2}\left( \xi \right) a-\nabla \left( \xi \right) \otimes
da+\nabla \left( \xi \right) \otimes da+\xi \otimes d^{2}a \\
&=&\nabla ^{2}\left( \xi \right) a\;,
\end{eqnarray*}

\noindent $\nabla ^{2}$ is $\mathcal{A}$-linear. And as $\mathcal{E}$ is a
projective $\mathcal{A}$-module, 
\[
\theta =\nabla ^{2}\in \limfunc{End}{}_{\mathcal{A}}\mathcal{E}\otimes _{%
\mathcal{A}}\Omega ^{2}\left( \mathcal{A}\right) =M\left( \mathcal{A}\right)
\otimes _{\mathcal{A}}\Omega ^{2}\left( \mathcal{A}\right) 
\]

\noindent is a matrix with elements in $\Omega ^{2}\left( \mathcal{A}\right) 
$.

Now, $\nabla $ defines a connection $\left[ \nabla ,\bullet \right] $ on $%
\func{End}_{\mathcal{A}}\mathcal{E}$ by 
\[
\begin{array}{cccc}
\left[ \nabla ,\bullet \right] : & \func{End}_{\mathcal{A}}\mathcal{E}%
\otimes _{\mathcal{A}}\Omega ^{p}\left( \mathcal{A}\right) & \rightarrow & 
\func{End}_{\mathcal{A}}\mathcal{E}\otimes _{\mathcal{A}}\Omega ^{p+1}\left( 
\mathcal{A}\right) \\ 
& \alpha & \mapsto & \left[ \nabla ,\alpha \right] =\nabla \circ \alpha
-\alpha \circ \nabla%
\end{array}
\]

\noindent and the curvature $\theta $ satisfies the \emph{Bianchi identity} $%
\left[ \nabla ,\theta \right] =0$.

On the other hand, a vector potential $A$ is a self-adjoint operator
interpreting a $1$-form 
\[
A=\sum_{j}a_{j}[D,b_{j}]
\]

\noindent and the associated force is the curvature $2$-form 
\[
\theta =dA+A^{2}~.
\]

The unitary group $\mathcal{U}(\mathcal{A})$ acts by gauge transformations
on $A$ and its $2$-form curvature $\theta $ 
\begin{eqnarray*}
A &\rightarrow &uAu^{\ast }+udu^{\ast }=uAu^{\ast }+u[D,u^{\ast }] \\
\theta  &\rightarrow &u\theta u^{\ast }~.
\end{eqnarray*}

\subsection{Axioms for geometry}

There are characteristic properties of classical (commutative) and
noncommutative geometries which can be used to axiomatize them.

$1$. (Classical and NC geometry). $ds=D^{-1}$ is an infinitesimal of order $%
\frac{1}{n}$ ($n$ is the dimension)~\footnote{%
In the NC\ framework, $ds$ and $dx$ are completely different sort of
entities.\ $dx$ is the differential of a coordinate and $ds$ doesn't commute
with it. In the classical case, the order of $ds$ as an infinitesimal is not 
$1$ but $1/n$. As we will see later, the Hilbert-Einstein action is the NC
integral of $ds^{n-2}$.} and for any $a\in \mathcal{A}$ integration is given
by $\limfunc{Tr}\nolimits_{Dix}\left( a\left\vert D\right\vert ^{-n}\right) $
(which is well defined and $\neq 0$ since $\left\vert D\right\vert ^{-n}$ is
an infinitesimal of order $1$). One can normalize the integral dividing by $%
V=\limfunc{Tr}\nolimits_{Dix}\left( \left\vert D\right\vert ^{-n}\right) $.

$2$. (Classical geometry). Universal commutation relations: $\left[ \left[
D,a\right] ,b\right] =0$, $\forall a,b\in \mathcal{A}$. So (Jones, Moscovici 
\cite{Jones})

\begin{quotation}
\noindent 
``while $ds$ no longer commutes with the coordinates, the algebra they
generate does satisfy non trivial commutation relations.''
\end{quotation}

$3$. (Classical and NC geometry). $a\in \mathcal{A}$ is ``smooth'' in the
sense that $a$ and $\left[ D,a\right] $ belong to the intersection of the
domains of the functionals $\delta ^{m}$ where $\delta \left( T\right) =%
\left[ \left| D\right| ,T\right] $ for every operator $T$ on $\mathcal{H}$.

$4$. (Classical geometry). If the dimension $n$ is \emph{even} there exists
a $\widetilde{\gamma }$ interpreting a $n$-form $c\in Z_{n}\left( \mathcal{A}%
,\mathcal{A}\right) $ associated to orientation and chirality (the $\gamma
^{5}$ of Dirac), $\widetilde{\gamma }$ being of the form $a_{0}\left[ D,a_{1}%
\right] \ldots \left[ D,a_{n}\right] $ and s.t. $\widetilde{\gamma }=%
\widetilde{\gamma }^{*}$ (self-adjointness), $\widetilde{\gamma }^{2}=1$, $%
\left\{ \widetilde{\gamma },D\right\} =0$ (anti-commutation relation) and $%
\left[ \widetilde{\gamma },a\right] =0$, $\forall a\in \mathcal{A}$
(commutation relations). $\widetilde{\gamma }$ decomposes $D$ into two parts 
$D=D^{+}+D^{^{\_}}$ where $D^{+}=(1-p)Dp$ with $p=\frac{1+\widetilde{\gamma }%
}{2}$. If $e$ is a self-adjoint ($e=e^{*}$) idempotent ($e^{2}=e$) of $%
\mathcal{A}$ (i.e. a projector), $eD^{+}e$ is a Fredholm operator from the
subspace $ep\mathcal{H}$ to the subspace $e(1-p)\mathcal{H}$. This can be
extended to the projectors of $e\in M_{q}\left( \mathcal{A}\right) $
defining finite projective left $\mathcal{A}$-modules $\mathcal{E}=\mathcal{A%
}^{N}e$ (if $\xi \in \mathcal{E}$ then $\xi e=\xi $) with the $\mathcal{A}$%
-valued inner product $\left( \xi ,\eta \right) =\sum_{i=1}^{i=N}\xi
_{i}\eta _{i}^{*}$.

$4$ bis.\ (Classical geometry). If $n$ is odd we ask only that there exists
such an $n$-form $c$ interpreted by $1$: $a_{0}\left[ D,a_{1}\right] \ldots %
\left[ D,a_{n}\right] =1$.

$5$. (Classical and NC geometry). $\mathcal{H}_{\infty }=\bigcap\limits_{m}%
\func{Domain}\left( D^{m}\right) $ is finite and projective as $\mathcal{A}$
-module and the formula $\left\langle a\xi ,\eta \right\rangle =\limfunc{Tr}%
_{Dix}a\left( \xi ,\eta \right) ds^{n}$ ($\left( \xi ,\eta \right) $ being
the scalar product of $\mathcal{H}$ and $\limfunc{Tr}_{Dix}$ the Dixmier
trace of infinitesimals of order $1$) defines an Hermitian structure on $%
\mathcal{H}_{\infty }$.

$6$. (Classical geometry). One can define an \emph{index pairing} of $D$
with $K_{0}\left( \mathcal{A}\right) $ and an \emph{intersection form} on $%
K_{0}\left( \mathcal{A}\right) $~\footnote{%
Remember that $K_{0}\left( \mathcal{A}\right) =\pi _{1}\left( GL_{\infty
}\left( \mathcal{A}\right) \right) $ classifies the finite projective $%
\mathcal{A}$-modules and that $K_{1}\left( \mathcal{A}\right) =\pi
_{0}\left( GL_{\infty }\left( \mathcal{A}\right) \right) $ is the group of
connected components of $GL_{\infty }\left( \mathcal{A}\right) $.}. If $%
\left[ \mathcal{E}\right] \in K_{0}\left( \mathcal{A}\right) $ is defined by
the projector $e$, we consider the scalar product $\left\langle \func{Ind}%
D,e\right\rangle $ which is an integer.\ We define therefore $\left\langle 
\func{Ind}D,e\right\rangle :K_{0}\left( \mathcal{A}\right) \rightarrow 
\mathbb{Z}$. As $\mathcal{A}$ is commutative, we can take the multiplication 
$m:\mathcal{A}\otimes \mathcal{A}\rightarrow \mathcal{A}$ given by $%
m(a\otimes b)=ab$ which induces $m_{0}:K_{0}\left( \mathcal{A}\right)
\otimes K_{0}\left( \mathcal{A}\right) \rightarrow K_{0}\left( \mathcal{A}%
\right) $.\ Composing with $\func{Ind}D$ we get the intersection form 
\begin{eqnarray*}
\left\langle \func{Ind}D,m_{0}\right\rangle  &:&K_{0}\left( \mathcal{A}%
\right) \otimes K_{0}\left( \mathcal{A}\right) \rightarrow \mathbb{Z} \\
(e,a) &\rightarrow &\left\langle \func{Ind}D,m_{0}(e\otimes a)\right\rangle
~.
\end{eqnarray*}

\emph{Poincar\'{e} duality}: the intersection form is invertible.

$7$. \emph{Real structure} (Classical geometry). There exists an anti-linear
isometry (charge conjugation) $J:\mathcal{H}\rightarrow \mathcal{H}$ which
combines charge conjugation and complex conjugation and gives the $*$%
-involution by algebraic conjugation: $JaJ^{-1}=a^{*}$ $\forall a\in 
\mathcal{A}$, and s.t. $J^{2}=\varepsilon $, $JD=\varepsilon ^{\prime }DJ$,
and $J\gamma =\varepsilon ^{\prime \prime }\gamma J$ with $\varepsilon $, $%
\varepsilon ^{\prime }$, $\varepsilon ^{\prime \prime }=\pm 1$ depending of
the dimension $n$ $\limfunc{mod}8$: 
\[
\begin{tabular}{|c|c|c|c|c|c|c|c|c|}
\hline
$n$ & $0$ & $1$ & $2$ & $3$ & $4$ & $5$ & $6$ & $7$ \\ \hline
$\varepsilon $ & $1$ & $1$ & $-1$ & $-1$ & $-1$ & $-1$ & $1$ & $1$ \\ \hline
$\varepsilon ^{\prime }$ & $1$ & $-1$ & $1$ & $1$ & $1$ & $-1$ & $1$ & $1$
\\ \hline
$\varepsilon ^{\prime \prime }$ & $1$ &  & $-1$ &  & $1$ &  & $-1$ &  \\ 
\hline
\end{tabular}
\]

In the classical case ($M$ smooth compact manifold of dimension $n$), Connes
proved that these axioms define a unique Riemannian spin geometry whose
geodesic distance and the spin structure are those defined by $D$.\
Moreover, the value of the Dixmier trace $\limfunc{Tr}_{Dix}ds^{n-2}$ is the 
\emph{Einstein-Hilbert action functional}: 
\[
\limfunc{Tr}\nolimits_{Dix}ds^{n-2}=c_{n}\int_{M}R\sqrt{g}%
d^{n}x=c_{n}\int_{M}Rdv 
\]

\noindent with $dv$ the volume form $dv=\sqrt{g}d^{n}x$ and $c_{n}=\frac{n-2%
}{12}\left( 4\pi \right) ^{-\frac{n}{2}}\Gamma \left( \frac{n}{2}+1\right)
^{-1}2^{\left[ \frac{n}{2}\right] }$.\ $\limfunc{Tr}\nolimits_{Dix}ds^{n-2}$
is well defined and $\neq 0$ since $ds^{n-2}$ is an infinitesimal of order $%
\frac{n-2}{n}<1$).\ For $n=4$, $c_{4}=\frac{1}{6}\left( 4\pi \right)
^{-2}\Gamma \left( 3\right) ^{-1}2^{2}=\frac{1}{48\pi ^{2}}$.

In the NC case the characteristic properties (2), (6), (7) must be modified
to take into account the NC:

$7^{NC}$. \emph{Real structure} (NC geometry). In the noncommutative case,
the axiom $JaJ^{-1}=a^{*}$ is transformed into the following axiom saying
that the conjugation by $J$ of the involution defines the \emph{opposed}
multiplication of $\mathcal{A}$.\ Let $b^{0}=Jb^{*}J^{-1}$, then $\left[
a,b^{0}\right] =0$ , $\forall a,b\in \mathcal{A}$. By means of this real
structure, the Hilbert space $\mathcal{H}$ becomes not only a (left) $%
\mathcal{A}$-module through the representation of $\mathcal{A}$ into $%
\mathcal{L}(\mathcal{H})$ but also a $\mathcal{A}\otimes \mathcal{A}^{\circ
} $-module (where $\mathcal{A}^{\circ }$ is the opposed algebra of $\mathcal{%
A})$ or a (left-right) $\mathcal{A}$-bimodule through $\left( a\otimes
b^{0}\right) \xi =aJb^{*}J^{-1}\xi $ or $a\xi b=aJb^{*}J^{-1}\xi $ for every 
$\xi \in \mathcal{H}$.

$2^{NC}$. The universal commutation relations $\left[ \left[ D,f\right] ,g%
\right] =0$, $\forall f,g\in \mathcal{A}$ become in the NC case $\left[ %
\left[ D,a\right] ,b^{\circ }\right] =0$, $\forall a,b\in \mathcal{A}$
(which is equivalent to $\left[ \left[ D,b^{\circ }\right] ,a\right] =0$, $%
\forall a,b\in \mathcal{A}$ since $a$ and $b^{\circ }$ commute by $7^{NC}$).

$6^{NC}$. $K$-theory can be easily generalized to the NC case.$\;$We
consider finite projective $\mathcal{A}$-modules $\mathcal{E}$, that is
direct factors of free $\mathcal{A}$-modules $\mathcal{A}^{N}$.$\;$They are
characterized by a projection $\pi :\mathcal{A}^{N}\rightarrow \mathcal{E}$
admitting a section $s:\mathcal{E}\rightarrow \mathcal{A}^{N}$ ($\pi \circ
s=Id_{\mathcal{E}}$). $K_{0}\left( \mathcal{A}\right) $ classifies them. The
structure of $\mathcal{A}\otimes \mathcal{A}^{\circ }$-module induced by the
real structure $J$ allows to define the intersection form by $%
(e,a)\rightarrow \left\langle \func{Ind}D,e\otimes a^{\circ }\right\rangle $
with $e\otimes a^{\circ }$ considered as an element of $K_{0}\left( \mathcal{%
\ A}\otimes \mathcal{A}^{\circ }\right) $.

One of the fundamental aspects of the NC case is that inner automorphisms $%
\alpha _{u}\left( a\right) =uau^{*}$, $u\in \mathcal{U}\left( \mathcal{A}%
\right) $ act upon the Dirac operator $D$ via NC gauge connections (vector
potentials) $A$%
\begin{eqnarray*}
\widetilde{D} &=&D+A+JAJ^{-1}\text{ with } \\
A &=&u\left[ D,u^{*}\right]
\end{eqnarray*}

\noindent the equivalence between $D$ and $\widetilde{D}$ being given by $%
\widetilde{D}=UDU^{-1}$ with $U=uJuJ^{-1}=u\left( u^{*}\right) ^{\circ }$.

\subsection{The crucial discovery of a structural link between ``external''
metric and ``internal'' gauge transformations}

One can generalize these transformations of metrics to gauge connections $A$
of the form $A=\sum a_{i}\left[ D,b_{i}\right] $ which can be interpreted as
internal perturbations of the metric or as \emph{internal fluctuations of
the spectral geometry} induced by the internal degrees of freedom of gauge
transformations. This coupling between metric and gauge transformations is
what is needed for coupling gravity with quantum field theory. In the
commutative case, this coupling vanishes since $U=uu^{\ast }=1$ and
therefore $\widetilde{D}=D$. The vanishing $A+JAJ^{-1}=0$ comes from the
fact that $A$ is self-adjoint and that, due to its special form $A=a\left[
D,b\right] $, we have $JAJ^{-1}=-A^{\ast }$. Indeed, since $\left[ D,b^{\ast
}\right] =-\left[ D,b\right] ^{\ast }$%
\begin{eqnarray*}
JAJ^{-1} &=&Ja\left[ D,b\right] J^{-1}=JaJ^{-1}J\left[ D,b\right]
J^{-1}=a^{\ast }\left[ D,b^{\ast }\right]  \\
&=&-a^{\ast }\left[ D,b\right] ^{\ast }=-\left( a\left[ D,b\right] \right)
^{\ast }=-A^{\ast }~.
\end{eqnarray*}

So the coupling between the \textquotedblleft external\textquotedblright\
metric afforded by the Dirac operator and the internal quantum degrees of
freedom is a purely noncommutative effect which constitutes a breakthrough
for the unification of general relativity and quantum field theory in a
\textquotedblleft good\textquotedblright\ theory of quantum gravity.

\subsection{Generating the Standard Model (Connes-Lott)}

Before concluding this compilation with some remarks on quantum gravity, let
us recall that the first main interest of noncommutative geometry in physics
was to couple classical gauge theories with purely NC such theories.\ This
led to the NC\ interpretation of Higgs fields. Connes' main result is:

\textbf{Connes' theorem.}\ The Glashow-Weinberg-Salam Standard Model (SM)
can be entirely reconstructed from the NC $C^{*}$-algebra 
\[
\mathcal{A}=\mathcal{C}^{\infty }(M)\otimes (\mathbb{C}\oplus \mathbb{H}%
\oplus M^{3}(\mathbb{C})) 
\]

\noindent where the \textquotedblleft internal\textquotedblright\ algebra $%
\mathbb{C}\oplus \mathbb{H}\oplus M^{3}(\mathbb{C})$ has for unitary group
the symmetry group 
\[
U(1)\times SU(2)\times SU(3)~.
\]

The first step is to construct the toy model which is the product $\mathcal{C%
}^{\infty }(M)\otimes (\mathbb{C}\oplus \mathbb{C})$ of the classical Dirac
fermionic model $\left( \mathcal{A}_{1},\mathcal{H}_{1},D_{1},\gamma
_{5}\right) $ and the previously explained, purely NC, $2$-points model $%
\left( \mathcal{A}_{2},\mathcal{H}_{2},D_{2},\gamma \right) $ with $D_{2}=%
\left[ 
\begin{array}{cc}
0 & M^{*} \\ 
M & 0%
\end{array}
\right] $: 
\[
\left\{ 
\begin{array}{l}
\mathcal{A}=\mathcal{A}_{1}\otimes \mathcal{A}_{2} \\ 
\mathcal{H}=\mathcal{H}_{1}\oplus \mathcal{H}_{2} \\ 
D=D_{1}\otimes 1+\gamma _{5}\otimes D_{2}\;.%
\end{array}
\right. 
\]

The second step is to complexify the model and to show that it enables to
derive the complete GWS Lagrangian.

The key idea is to take the product of a $4$-dimensional spin manifold $M$
with a finite NC geometry $\left( \mathcal{A}_{F},\mathcal{H}%
_{F},D_{F}\right) $ of dimension $0$ where $\mathcal{H}_{F}$ is the Hilbert
space with basis the generations of fermions: quarks and leptons. The
particule/antiparticule duality decomposes $\mathcal{H}_{F}$ into $\mathcal{H%
}_{F}=\mathcal{H}_{F}^{+}\oplus \mathcal{H}_{F}^{-}$, each $\mathcal{H}%
_{F}^{\pm }$ decomposes into $\mathcal{H}_{F}^{\pm }=\mathcal{H}_{l}^{\pm
}\oplus \mathcal{H}_{q}^{\pm }$ ($l=$ lepton and $q=$ quark), and chirality
decomposes the $\mathcal{H}_{p}^{\pm }$ ($p=$ particule) into $\mathcal{H}%
_{pL}^{\pm }\oplus \mathcal{H}_{pR}^{\pm }$ ($L=$ left, $R=$ right). The $4$
quarks are $u_{L},u_{R},d_{L},d_{R}$ ($u=$ up, $d=$ down) with $3$ colours ($%
12$ quarks for each generation) and the $3$ leptons are $e_{L},\nu _{L},e_{R}
$, the total being of $2\left( 12+3\right) =30$ fermions for each generation.

The real structure $J$ is given for $\mathcal{H}_{F}=\mathcal{H}%
_{F}^{+}\oplus \mathcal{H}_{F}^{-}$ by $J\left( 
\begin{array}{c}
\xi  \\ 
\overline{\eta }%
\end{array}%
\right) =\left( 
\begin{array}{c}
\eta  \\ 
\overline{\xi }%
\end{array}%
\right) $ that is, if $\xi =\sum_{i}\lambda _{i}p_{i}$ and $\overline{\eta }%
=\sum_{j}\mu _{j}\overline{p_{j}}$, 
\[
J\left( \sum_{i}\lambda _{i}p_{i}+\sum_{j}\mu _{j}\overline{p_{j}}\right)
=\left( \sum_{j}\overline{\mu _{j}}p_{j}+\sum_{i}\overline{\lambda _{i}}%
\overline{p_{i}}\right) ~.
\]

The action of the internal algebra $\mathcal{A}_{F}=\mathbb{C}\oplus \mathbb{%
H}\oplus M^{3}(\mathbb{C})$ is defined in the following way.\ Let $a=\left(
\lambda ,q,m\right) \in \mathcal{A}_{F}$, $\lambda \in \mathbb{C}$ being a
complex scalar acting upon $\mathbb{C}^{2}$ as the diagonal quaternion $%
\left( 
\begin{array}{cc}
\lambda  & 0 \\ 
0 & \overline{\lambda }%
\end{array}%
\right) $, $q=\alpha +\beta j\in \mathbb{H}$ being a quaternion written as $%
\left( 
\begin{array}{cc}
\alpha  & \beta  \\ 
-\overline{\beta } & \overline{\alpha }%
\end{array}%
\right) $ with $j=\left( 
\begin{array}{cc}
0 & 1 \\ 
-1 & 0%
\end{array}%
\right) $, and $m\in M^{3}(\mathbb{C})$ being a $3\times 3$ complex matrix.
The element $a=\left( \lambda ,q,m\right) $ acts on quarks, independently of
color, via $au_{R}=\lambda u_{R}$, $au_{L}=\alpha u_{L}-\overline{\beta }%
d_{L}$, $ad_{R}=\overline{\lambda }d_{R}$, $ad_{L}=\beta u_{L}+\overline{%
\alpha }d_{L}$, that is as

\[
\left( \lambda ,q,m\right) \left( 
\begin{array}{c}
u_{L} \\ 
d_{L} \\ 
u_{R} \\ 
d_{R}%
\end{array}
\right) =\left( 
\begin{array}{cccc}
\alpha & -\overline{\beta } & 0 & 0 \\ 
\beta & \overline{\alpha } & 0 & 0 \\ 
0 & 0 & \lambda & 0 \\ 
0 & 0 & 0 & \overline{\lambda }%
\end{array}
\right) \left( 
\begin{array}{c}
u_{L} \\ 
d_{L} \\ 
u_{R} \\ 
d_{R}%
\end{array}
\right) =\left( 
\begin{array}{c}
\alpha u_{L}-\overline{\beta }d_{L} \\ 
\beta u_{L}+\overline{\alpha }d_{L} \\ 
\lambda u_{R} \\ 
\overline{\lambda }d_{R}%
\end{array}
\right) 
\]

\noindent (the pair $\left( u_{R},d_{R}\right) $ can be considered as an
element of $\mathbb{C}\oplus \mathbb{C}$, while $\left( u_{L},d_{L}\right) $
can be considered as an element of $\mathbb{C}^{2}$). It acts on leptons via 
$ae_{R}=\overline{\lambda }e_{R}$, $ae_{L}=\beta \nu _{L}+\overline{\alpha }%
e_{L}$, $a\nu _{L}=\alpha \nu _{L}-\overline{\beta }e_{L}$, that is as 
\[
\left( \lambda ,q,m\right) \left( 
\begin{array}{c}
e_{R} \\ 
\nu _{L} \\ 
e_{L}%
\end{array}%
\right) =\left( 
\begin{array}{ccc}
\overline{\lambda } & 0 & 0 \\ 
0 & \alpha  & -\overline{\beta } \\ 
0 & \beta  & \overline{\alpha }%
\end{array}%
\right) \left( 
\begin{array}{c}
e_{R} \\ 
\nu _{L} \\ 
e_{L}%
\end{array}%
\right) =\left( 
\begin{array}{c}
\overline{\lambda }e_{R} \\ 
\alpha \nu _{L}-\overline{\beta }e_{L} \\ 
\beta \nu _{L}+\overline{\alpha }e_{L}%
\end{array}%
\right) ~.
\]

\noindent It acts on anti-particules via $a\overline{l}=\lambda \overline{l}$
for antileptons and via $a\overline{q}=m\overline{q}$ for antiquarks where $%
m $ acts upon color.

The internal Dirac operator $D_{F}$ is given by the matrix of Yukawa
coupling $D_{F}=\left( 
\begin{array}{cc}
Y & 0 \\ 
0 & \overline{Y}%
\end{array}
\right) $ where $Y=\left( Y_{q}\otimes 1_{3}\right) \oplus Y_{l}$ (the $%
\otimes 1_{3}$ comes from the $3$ generations of fermions) with 
\[
Y_{q}= 
\begin{array}{cc}
& 
\begin{array}{cccc}
u_{L}\; & d_{L}\; & u_{R}\; & d_{R}\;%
\end{array}
\; \\ 
\begin{array}{c}
u_{L} \\ 
d_{L} \\ 
u_{R} \\ 
d_{R}%
\end{array}
& \left( 
\begin{array}{cccc}
0 & 0 & M_{u} & 0 \\ 
0 & 0 & 0 & M_{d} \\ 
M_{u}^{*} & 0 & 0 & 0 \\ 
0 & M_{d}^{*} & 0 & 0%
\end{array}
\right)%
\end{array}
\]

\noindent and 
\[
Y_{l}= 
\begin{array}{cc}
& 
\begin{array}{ccc}
e_{R} & \nu _{L} & e_{L}%
\end{array}
\\ 
\begin{array}{c}
e_{R} \\ 
\nu _{L} \\ 
e_{L}%
\end{array}
& \left( 
\begin{array}{ccc}
0 & 0 & M_{l} \\ 
0 & 0 & 0 \\ 
M_{l}^{*} & 0 & 0%
\end{array}
\right)%
\end{array}
\]

\noindent where (Connes \cite{Connes96}) $M_{u}$, $M_{d}$, and $M_{l}$ are
matrices \textquotedblleft which encode both the masses of the Fermions and
their mixing properties\textquotedblright . Chirality is given by $\gamma
_{F}\left( p_{R}\right) =p_{R}$ and $\gamma _{F}\left( p_{L}\right) =-p_{L}$
($p$ being any particule or anti-particule).

Connes and Lott then take the product of this internal model of the
fermionic sector with a classical gauge model for the bosonic sector: 
\[
\left\{ 
\begin{array}{l}
\mathcal{A}=C^{\infty }\left( M\right) \otimes \mathcal{A}_{F}=\left(
C^{\infty }\left( M\right) \otimes \mathbb{C}\right) \oplus \left( C^{\infty
}\left( M\right) \otimes \mathbb{H}\right) \oplus \left( C^{\infty }\left(
M\right) \otimes M^{3}(\mathbb{C})\right)  \\ 
\mathcal{H}=L^{2}\left( M,S\right) \otimes \mathcal{H}_{F}=L^{2}\left(
M,S\otimes \mathcal{H}_{F}\right)  \\ 
D=\left( D_{M}\otimes 1\right) \oplus \left( \gamma _{5}\otimes D_{F}\right)
~.%
\end{array}%
\right. 
\]

The extraordinary \textquotedblleft tour de force\textquotedblright\ is that
this model, which is rather simple at the conceptual level (a product of two
models, respectively fermionic and bosonic, which takes into account only
the known fundamental properties of these two sectors), is in fact extremely
complex and generates the standard model in a \emph{principled} way.
Computations are very intricate (see Kastler's papers in the bibliography).\
One has to compute first vector potentials of the form $A=\sum_{i}a_{i}\left[
D,a_{i}^{\prime }\right] $, $a_{i},a_{i}^{\prime }\in \mathcal{A}$ which
induce fluctuations of the metric.\ As $D$ is a sum of two terms, it is also
the case for $A$.\ Its discrete part comes from $\gamma _{5}\otimes D_{F}$
and generates the Higgs bosons.\ Let $a_{i}\left( x\right) =\left( \lambda
_{i}\left( x\right) ,q_{i}\left( x\right) ,m_{i}\left( x\right) \right) $.\
The term $\sum_{i}a_{i}\left[ \gamma _{5}\otimes D_{F},a_{i}^{\prime }\right]
$ yields $\gamma _{5}$ tensored by matrices of the form

\begin{itemize}
\item for the quark sector:
\end{itemize}

\[
\left( 
\begin{array}{cccc}
0 & 0 & M_{u}\varphi _{1} & M_{u}\varphi _{2} \\ 
0 & 0 & -M_{d}\overline{\varphi _{2}} & M_{d}\overline{\varphi _{1}} \\ 
M_{u}^{*}\varphi _{1}^{\prime } & M_{d}^{*}\varphi _{2}^{\prime } & 0 & 0 \\ 
-M_{u}^{*}\overline{\varphi _{2}^{\prime }} & M_{d}^{*}\overline{\varphi
_{1}^{\prime }} & 0 & 0%
\end{array}
\right) 
\]

with 
\[
\left\{ 
\begin{array}{l}
\varphi _{1}=\sum_{i}\lambda _{i}\left( \alpha _{i}^{\prime }-\lambda
_{i}^{\prime }\right) \\ 
\varphi _{2}=\sum_{i}\lambda _{i}\beta _{i}^{\prime } \\ 
\varphi _{1}^{\prime }=\sum_{i}\alpha _{i}\left( \lambda _{i}^{\prime
}-\alpha _{i}^{\prime }\right) +\beta _{i}\overline{\beta _{i}^{\prime }} \\ 
\varphi _{2}^{\prime }=\sum_{i}\beta _{i}\left( \overline{\lambda
_{i}^{\prime }}-\overline{\alpha _{i}^{\prime }}\right) -\alpha _{i}\beta
_{i}^{\prime }\;.%
\end{array}
\right. 
\]

\begin{itemize}
\item and for the lepton sector: 
\[
\left( 
\begin{array}{ccc}
0 & -M_{d}\overline{\varphi _{2}} & M_{d}\overline{\varphi _{1}} \\ 
M_{d}^{\ast }\varphi _{2}^{\prime } & 0 & 0 \\ 
M_{d}^{\ast }\overline{\varphi _{1}^{\prime }} & 0 & 0%
\end{array}%
\right) ~.
\]
\end{itemize}

Let $q=\varphi _{1}+\varphi _{2}j$ and $q^{\prime }=\varphi _{1}^{\prime
}+\varphi _{2}^{\prime }j$ be the quaternionic fields so defined. As $%
A=A^{*} $, we have $q^{\prime }=q^{*}$.\ The $\mathbb{H}$-valued field $q(x)$
is the \emph{Higgs doublet}.

The second part of the vector potential $A$ comes from $D_{M}\otimes 1$ and
generates the gauge bosons. The terms $\sum_{i}a_{i}\left[ D_{M}\otimes
1,a_{i}^{\prime }\right] $ yield

\begin{itemize}
\item the $U(1)$ gauge field $\Lambda =\sum_{i}\lambda _{i}d\lambda
_{i}^{\prime }$;

\item the $SU(2)$ gauge field $Q=\sum_{i}q_{i}dq_{i}^{\prime }$;

\item the $U(3)$ gauge field $V=\sum_{i}m_{i}dm_{i}^{\prime }$.
\end{itemize}

The computation of the fluctuations of the metric $A+JAJ^{-1}$ gives

\begin{itemize}
\item for the quark sector:
\end{itemize}

\[
\begin{array}{cc}
& 
\begin{array}{cccc}
u_{L}\;\;\;\;\; & d_{L}\;\;\;\;\; & u_{R}\;\;\;\;\; & d_{R}\;\;\;\;\;%
\end{array}
\\ 
\begin{array}{c}
u_{L} \\ 
d_{L} \\ 
u_{R} \\ 
d_{R}%
\end{array}
& \left( 
\begin{array}{cccc}
Q_{11}1_{3}+V & Q_{12}1_{3} & 0 & 0 \\ 
Q_{21}1_{3} & Q_{22}1_{3}+V & 0 & 0 \\ 
0 & 0 & \Lambda 1_{3}+V & 0 \\ 
0 & 0 & 0 & -\Lambda 1_{3}+V%
\end{array}
\right)%
\end{array}
\]

\noindent which is a $12\times 12$ matrix since $V$ is $3\times 3$,

\begin{itemize}
\item and for the lepton sector: 
\[
\begin{array}{cc}
& 
\begin{array}{ccc}
e_{R}\;\;\;\;\; & \nu _{L}\;\;\;\;\; & e_{L}\;\;\;\;\;%
\end{array}
\\ 
\begin{array}{c}
e_{R} \\ 
\nu _{L} \\ 
e_{L}%
\end{array}
& \left( 
\begin{array}{ccc}
-2\Lambda  & 0 & 0 \\ 
0 & Q_{11}-\Lambda  & Q_{12} \\ 
0 & Q_{21} & Q_{22}-\Lambda 
\end{array}%
\right) 
\end{array}%
~.
\]
\end{itemize}

\noindent One can suppose moreover that $\func{Trace}V=\Lambda $, that is $%
V=V^{\prime }+\frac{1}{3}\Lambda $ with $V^{\prime }$ traceless, which gives
the correct hypercharges.

The crowning of the computation is that the total (bosonic$~+$ fermionic)
action 
\[
\limfunc{Tr}\nolimits_{Dix}\theta ^{2}ds^{4}+\left\langle \left(
D+A+JAJ^{-1}\right) \psi ,\psi \right\rangle =YM\left( A\right)
+\left\langle D_{A}\psi ,\psi \right\rangle 
\]

\noindent (where $\theta =dA+A^{2}$ is the curvature of the connection $A$)
enables to derive the complete GWS Lagrangian

\[
\mathcal{L}=\mathcal{L}_{G}+\mathcal{L}_{f}+\mathcal{L}_{\varphi }+\mathcal{L%
}_{Y}+\mathcal{L}_{V}\;. 
\]

1. $\mathcal{L}_{G}$ is the Lagrangian of the gauge bosons 
\begin{eqnarray*}
\mathcal{L}_{G} &=&\frac{1}{4}\left( G_{\mu \nu a}G_{a}^{\mu \nu }\right) +%
\frac{1}{4}\left( F_{\mu \nu }F^{\mu \nu }\right)  \\
G_{\mu \nu a} &=&\partial _{\mu }W_{\nu a}-\partial _{\nu }W_{\mu
a}+g\varepsilon _{abc}W_{\mu b}W_{\nu c}\text{, } \\
&&\text{where }W_{\mu a}\text{ is a }SU(2)\text{ gauge field (weak isospin)}
\\
F_{\mu \nu } &=&\partial _{\mu }B_{\nu }-\partial _{\nu }B_{\mu }\text{,
with }B_{\mu }\text{ a }SU(1)\text{ gauge field. }
\end{eqnarray*}

2.$\;\mathcal{L}_{f}$ is the fermionic kinetic term 
\begin{eqnarray*}
\mathcal{L}_{f} &=&-\sum \overline{f_{L}}\gamma ^{\mu }\left( \partial _{\mu
}+ig\frac{\tau _{a}}{2}W_{\mu a}+ig^{\prime }\frac{Y_{L}}{2}B_{\mu }\right)
f_{L}+ \\
&&\overline{f_{R}}\gamma ^{\mu }\left( \partial _{\mu }+ig^{\prime }\frac{%
Y_{R}}{2}B_{\mu }\right) f_{R}
\end{eqnarray*}

\noindent where $f_{L}=\left[ 
\begin{array}{l}
\nu _{L} \\ 
e_{L}%
\end{array}%
\right] $ are left fermion fields of hypercharge $Y_{L}=-1$ and $%
f_{R}=\left( e_{R}\right) $ right fermion fields of hypercharge $Y_{R}=-2$.

3.$\;\mathcal{L}_{\varphi }$ is the Higgs kinetic term

\[
\mathcal{L}_{\varphi }=-\left| \left( \partial _{\mu }+ig\frac{\tau _{a}}{2}%
W_{\mu a}+i\frac{g^{\prime }}{2}B_{\mu }\right) \varphi \right| ^{2} 
\]

\noindent where $\varphi =\left[ 
\begin{array}{l}
\varphi _{1} \\ 
\varphi _{2}%
\end{array}%
\right] $ is a $SU(2)$ pair of scalar complex fields of hypercharge $%
Y_{\varphi }=1$.

4.$\;\mathcal{L}_{Y}$ is a Yukawa coupling between the Higgs fields and the
fermions 
\[
\mathcal{L}_{Y}=-\sum \left( H_{ff^{\prime }}\left( \overline{f_{L}}.\varphi
\right) f_{R}^{\prime }+H_{ff^{\prime }}^{*}\overline{f_{^{\prime }R}}\left(
\varphi ^{+}.f_{L}\right) \right) 
\]

\noindent where $H_{ff^{\prime }}$ is a coupling matrix.

5. $\mathcal{L}_{V}$ is the Lagrangian of the self-interaction of the Higgs
fields 
\[
\mathcal{L}_{V}=\mu ^{2}\left( \varphi ^{+}\varphi \right) -\frac{1}{2}%
\lambda \left( \varphi ^{+}\varphi \right) ^{2}\text{ with }\lambda >0~.
\]

\section{Quantum gravity, fluctuating background geometry, and spectral
invariance (Connes~- Chamseddine)}

\subsection{Quantum Field Theory and General Relativity}

As we have already emphasized, Alain Connes realized a new breakthrough in
the approaches of quantum gravity by coupling such models with general
relativity. In NCG, quantum gravity can be thought of in a principled way
because it becomes possible to introduce in the models of quantum field
theory the gravitational Einstein-Hilbert action as a direct consequence of
the specific invariance of spectral geometry, namely \emph{spectral
invariance}. As Alain Connes \cite{Connes96} explains:

\begin{quotation}
\noindent 
``However this [the previous NC deduction of the standard
model] requires the definition of the curvature and is still in the spirit
of gauge theories.\ (...) One should consider the internal gauge symmetries
as part of the diffeomorphism group of the non commutative geometry, and the
gauge bosons as the internal fluctuations of the metric.\ It follows then
that the action functional should be of a purely gravitational nature.\ We
state the principle of spectral invariance, stronger than the invariance
under diffeomorphisms, which requires that the action functional only
depends on the spectral properties of $D=ds^{-1}$ in $\mathcal{H}$%
.'' 
\end{quotation}

The general strategy for coupling a Yang-Mills-Higgs gauge theory with the
Einstein-Hilbert action is to find a $C^{*}$-algebra $\mathcal{A}$ s.t. the
normal subgroup $\func{Inn}(\mathcal{A})$ of inner automorphisms is the
gauge group and the quotient group $\func{Out}(\mathcal{A})=\func{Aut}(%
\mathcal{A})/\func{Inn}(\mathcal{A})$ of ``external'' automorphisms plays
the role of $\func{Diff}(M)$ in a gravitational theory. Indeed, in the
classical setting we have principal bundles $P\rightarrow M$ with a
structural group $G$ acting upon the fibers and an exact sequence 
\[
Id\rightarrow \mathcal{G}\rightarrow \func{Aut}\left( P\right) \rightarrow 
\func{Diff}(M)\rightarrow Id 
\]

\noindent where $\mathcal{G}=C^{\infty }\left( M,G\right) $ is the gauge
group.\ The non abelian character of these gauge theories comes solely\ from
the non commutativity of the group of \emph{internal} symmetries $G$. The
total symmetry group $\func{Aut}\left( P\right) $ of the theory is the
semidirect product $\mathfrak{G}$ of $\func{Diff}(M)$ and $\mathcal{G}%
=C^{\infty }\left( M,G\right) $.\ If we want to geometrize the theory
completely, we would have to find a generalized space $X$ s.t. $\func{Aut}%
\left( X\right) =\mathfrak{G}$.

\begin{quotation}
\noindent 
``If such a space would exist, then we would have some
chance to actually geometrize completely the theory, namely to be able to
say that it's pure gravity on the space $X$.'' (Connes \cite%
{Connes2000a})
\end{quotation}

But this is\emph{\ impossible }if $X$ is a\emph{\ }manifold since a theorem
of John Mather proves that in that case the group $\func{Diff}(X)$ would be
simple (without normal subgroup) and could'nt therefore be a semidirect
product. But it is possible with a NC\ space $\left( \mathcal{A},\mathcal{H}%
,D\right) $. For then (Iochum, Kastler, Sch\"{u}cker \cite{Iochum})

\begin{quotation}
\noindent 
``the metric `fluctuates', that is, it picks up additional degrees of
freedom from the internal space, the Yang-Mills connection and the Higgs
scalar. (...) In physicist's language, the spectral triplet is the Dirac
action of a multiplet of dynamical fermions in a background field.\ This
background field is a fluctuating metric, consisting of so far adynamical
bosons of spin $0$,$1$ and $2$''.
\end{quotation}

If we find a NC\ geometry $\mathcal{A}$ with $\func{Inn}(\mathcal{A})\simeq 
\mathcal{G}$, a correct spectral triple and apply the spectral action, then
gravity will correspond to $\func{Out}(\mathcal{A})=\func{Aut}(\mathcal{A})/%
\func{Inn}(\mathcal{A})$. As was emphasized by Martin \emph{et al}. \cite%
{Martin}:

\begin{quotation}
\noindent 
``The strength of Connes' conception is that gauge theories are thereby
deeply connected to the underlying geometry, on the same footing as
gravity.\ The distinction between gravitational and gauge theories boils
down to the difference between outer and inner automorphisms.''
\end{quotation}

\noindent Jones and Moscovici \cite{Jones} add that this implies that

\begin{quotation}
\noindent 
``Connes' spectral approach gains the ability to reach below the Planck
scale and attempt to decipher the fine structure of space-time''.
\end{quotation}

\noindent So, just as general relativity extends the Galilean or Minkowskian
invariance into diffeomorphism invariance, NCG extends both diffeomorphism
invariance and gauge invariance into a larger invariance, the spectral
invariance.

\subsection{The spectral action and the eigenvalues of the Dirac operator as
dynamical variables for general relativity}

The key device is the bosonic spectral action 
\[
\func{Trace}\left( \phi \left( \frac{D^{2}}{\Lambda ^{2}}\right) \right) 
\]

\noindent where $\Lambda $ is a cut-off of the order of the inverse of
Planck length and $\phi $ a smooth approximation of the characteristic
function $\chi _{\left[ 0,1\right] }$ of the unit interval. $D^{2}=\left(
D_{M}\otimes 1+\gamma _{5}\otimes D_{F}\right) ^{2}$ is computed using
Lichnerowicz' formula $D^{2}=\Delta ^{S}+\frac{1}{4}R$. As this action
counts the number $N\left( \Lambda \right) $ of eigenvalues of $D$ in the
interval $\left[ -\Lambda ,\Lambda \right] $, the key idea is, as formulated
by Giovanni Landi and Carlo Rovelli \cite{Landi},

\begin{quotation}
\noindent 
``to consider the eigenvalues of the Dirac operator as dynamical variables
for general relativity''.
\end{quotation}

This formulation highlights the physical and philosophical significance of
the NC framework: since the distance is defined through the Dirac operator $D
$, the spectral properties of $D$ can be used in order to modify the metric.
The eigenvalues are spectral invariants and are therefore, in the classical
case, automatically $\func{Diff}(M)$ invariant.\ 

\begin{quotation}
\noindent 
``Thus the general idea is to describe spacetime geometry by
giving the eigen-frequencies of the spinors that can live on that spacetime.
[...] The Dirac operator $D$ encodes the full information about the
spacetime geometry in a way usable for describing gravitational
dynamics.'' (Landi-Rovelli \cite{Landi}: the quotation
concerns our $D_{M}$ acting on the Hilbert space of spinor fields on $M$.)
\end{quotation}

This crucial point has also been well explained by Steven Carlip (\cite%
{Carlip}, p.\ 47).\ Due to $\func{Diff}(M)$ invariance, in general
relativity points of space-time loose any physical meaning so that
obervables must be radically non-local.\ This is the case with the
eingenvalues of $D$ which

\begin{quotation}
\noindent 
``provide a nice set of non local, diffeomorphism-invariant obervables.''
\end{quotation}

\noindent They yield

\begin{quotation}
\noindent 
``the first good candidates for a (nearly) complete set of
diffeomorphism-invariant observables''.
\end{quotation}

Let us look at $N\left( \Lambda \right) $ for $\Lambda \rightarrow \infty $%
.\ $N\left( \Lambda \right) $ is a step function which encodes a lot of
information and can be written as a sum of a mean value and a fluctuation
(oscillatory) term $N\left( \Lambda \right) =\left\langle N\left( \Lambda
\right) \right\rangle +N_{\func{osc}}\left( \Lambda \right) $ where the
oscillatory part $N_{\func{osc}}\left( \Lambda \right) $ is random. The mean
part $\left\langle N\left( \Lambda \right) \right\rangle $ can be computed
using a semi-classical approximation and a heat equation expansion. A
wonderful computation shows that for $n=4$ the asymptotic expansion of the
spectral action is 
\[
\func{Trace}\left( \phi \left( \frac{D^{2}}{\Lambda ^{2}}\right) \right)
=\Lambda ^{4}f_{0}a_{0}\left( D^{2}\right) +\Lambda ^{2}f_{2}a_{2}\left(
D^{2}\right) +f_{4}a_{4}\left( D^{2}\right) +O\left( \Lambda ^{-2}\right) 
\]

\noindent with

\begin{itemize}
\item $f_{0}=\int_{\mathbb{R}}\phi \left( u\right) udu$, $f_{2}=\int_{%
\mathbb{R}}\phi \left( u\right) du$, $f_{4}=\phi \left( 0\right) $.

\item $a_{j}\left( D^{2}\right) =\int_{M}a_{j}\left( x,D^{2}\right) dv$ ( $%
dv=\sqrt{g}d^{4}x$).

\item $a_{0}\left( x,D^{2}\right) =\frac{1}{\left( 4\pi \right) ^{2}}\func{%
Trace}_{x}\left( 1\right) $.

\item $a_{2}\left( x,D^{2}\right) =\frac{1}{\left( 4\pi \right) ^{2}}\func{%
Trace}_{x}\left( \frac{1}{6}s1-E\right) $.

\item $a_{4}\left( x,D^{2}\right) =\frac{1}{360\left( 4\pi \right) ^{2}}%
\func{Trace}_{x}\left( 5s^{2}1-2r^{2}1+2R^{2}1-60sE+180E^{2}+30R_{\mu \nu
}^{\nabla }R^{\nabla \mu \nu }\right) $.

\item $R$ is the curvature tensor of $M$ and $R^{2}=R_{\mu \nu \alpha \beta
}R^{\mu \nu \alpha \beta }$.

\item $r$ is the Ricci tensor of $M$ and $r^{2}=r_{\mu \nu }r^{\mu \nu }$.

\item $s$ is the scalar curvature of $M$.

\item $E$ and $R_{\mu \nu }^{\nabla }$ come from Lichnerowicz' formula.
\end{itemize}

Let 
\[
\mathcal{E}=C^{\infty }(M,S\otimes \mathcal{H}_{F})=C^{\infty }(M,S)\otimes
_{C^{\infty }(M)}C^{\infty }(M,\mathcal{H}_{F})~.\ 
\]

\noindent The connection on $\mathcal{E}$ is 
\[
\nabla =\nabla ^{S}\otimes Id_{C^{\infty }(M,\mathcal{H}_{F})}+Id_{C^{\infty
}(M,S)}\otimes \nabla ^{F} 
\]

\noindent and $R_{\mu \nu }^{\nabla }$ is the curvature $2$-tensor of this
total connection $\nabla $. If $D=ic^{\mu }\nabla _{\mu }+\varphi $ with $%
c^{\mu }=\gamma ^{\mu }\otimes Id_{C^{\infty }(M,\mathcal{H}_{F})}$, then $%
D^{2}=\Delta +E$, with

\[
\left\{ 
\begin{array}{l}
\Delta =-g^{\mu \nu }\left( \nabla _{\mu }\nabla _{\nu }-\Gamma _{\mu \nu
}^{\alpha }\nabla _{\alpha }\right)  \\ 
E=\frac{1}{4}s1-\frac{1}{2}c\left( R^{F}\right) +ic^{\mu }\left[ \nabla
_{\mu },\varphi \right] +\varphi ^{2} \\ 
c\left( R^{F}\right) =-\gamma ^{\mu }\gamma ^{\nu }\otimes R_{\mu \nu }^{F}%
\text{ (}R^{F}=\text{curvature of }\nabla ^{F}\text{)~.}%
\end{array}%
\right. 
\]

The asymptotic expansion of the spectral action is dominated by the first
two terms which can be identified with the Einstein-Hilbert action with a
cosmological term.\ The later can be eliminated by a change of $\phi $.

\textbf{Addendum}. In a forthcoming book, Alain Connes, Ali Chamseddine and
Matilde Marcolli show how the previous results can be strongly improved and
yield a derivation of the standard model minimally coupled to gravity
(Einstein-Hilbert action) with massive neutrinos, neutrino mixing, Weinberg
angle, and Higgs mass (of the order of 170 GeV).\ This new achievement is
quite astonishing.

\end{document}